\documentclass[pss]{wiley2sp} 
\usepackage{amsmath}

\newcommand{\BFCA}{\ensuremath{\mathrm{Ba(Fe}_{1-x}\mathrm{Co}_{x})_{2}\mathrm{As}_{2}}}

\newcommand{\BFCMA}{\ensuremath{\mathrm{Ba(Fe}_{1-x-y}\mathrm{Co}_{x}\mathrm{Mn}_{y})_{2}\mathrm{As}_{2}}}
\newcommand{\BFAP}{\ensuremath{\mathrm{BaFe}_{2}\mathrm{As}_{2-x}\mathrm{P}_{x}}}
\newcommand{\EFAP}{\ensuremath{\mathrm{EuFe}_{2}\mathrm{As}_{2-x}\mathrm{P}_{x}}}
\newcommand{\EFA}{\ensuremath{\mathrm{EuFe}_{2}\mathrm{As}_{2}}}
\newcommand{\BFA}{\ensuremath{\mathrm{BaFe}_{2}\mathrm{As}_{2}}}
\newcommand{\BCA}{\ensuremath{\mathrm{BaCo}_{2}\mathrm{As}_{2}}}
\newcommand{\BFP}{\ensuremath{\mathrm{BaFe}_{2}\mathrm{P}_{2}}}
\newcommand{\KBFA}{\ensuremath{\mathrm{K}_{x}\mathrm{Ba}_{1-x}\mathrm{Fe}_{2}\mathrm{As}_{2}}}

\newcommand{\NFCA}{\ensuremath{\mathrm{NaFe}_{1-x}\mathrm{Co}_{x}\mathrm{As}}}
\newcommand{\NFRA}{\ensuremath{\mathrm{NaFe}_{1-x}\mathrm{Rh}_{x}\mathrm{As}}}

\newcommand{\BSCCO}{\ensuremath{\mathrm{Bi}_2\mathrm{Sr}_2\mathrm{CaCu}_2\mathrm{O}_{8+\delta}}}

\begin{document}

\title{Electronic structure and ultrafast dynamics of FeAs-based superconductors by angle- and time-resolved photoemission
spectroscopy.}

\titlerunning{Electronic structure and ultrafast dynamics }

\author{%
I. Avigo\textsuperscript{\textsf{\bfseries 1}}, S.
Thirupathaiah\textsuperscript{\textsf{\bfseries 2}}, E. D. L.
Rienks\textsuperscript{\textsf{\bfseries 3,4}}, L.
Rettig\textsuperscript{\textsf{\bfseries 1}}, A.
Charnukha\textsuperscript{\textsf{\bfseries 5}}, M.
Ligges\textsuperscript{\textsf{\bfseries 1}}, R.
Cortes\textsuperscript{\textsf{\bfseries 6}}, J.
Nayak\textsuperscript{\textsf{\bfseries 7}}, H. S.
Jeevan\textsuperscript{\textsf{\bfseries 8}}, T.
Wolf\textsuperscript{\textsf{\bfseries 9}}, Y.
Huang\textsuperscript{\textsf{\bfseries 10}}, S.
Wurmehl\textsuperscript{\textsf{\bfseries 3}}, P.
Gegenwart\textsuperscript{\textsf{\bfseries 8}}, M. S.
Golden\textsuperscript{\textsf{\bfseries 10}}, B.
B\"uchner\textsuperscript{\textsf{\bfseries 3,4}}, M.
Vojta\textsuperscript{\textsf{\bfseries 4}}, M.
Wolf\textsuperscript{\textsf{\bfseries 6}}, C.
Felser\textsuperscript{\textsf{\bfseries 7}}, J.
Fink\textsuperscript{\Ast,\textsf{\bfseries 3,4,7}}, U.
Bovensiepen\textsuperscript{\Ast,\textsf{\bfseries 1}} }

\authorrunning{I. Avigo et al.}

\mail{e-mail \textsf{uwe.bovensiepen@uni-due.de}, Phone
+49-203-3794566, Fax +49-203-3794555;\\
\textsf{J.Fink@ifw-dresden.de}, Phone
  +49-351-4659425, Fax +49-351-4659313}

\institute{%
 \textsuperscript{1}\,Fakult\"at f\"ur Physik, Universit\"at Duisburg-Essen, Lotharstr. 1, D-47057 Duisburg, Germany.\\
  \textsuperscript{2}\,Solid State and Structural Chemistry Unit, Indian Institute of Science, Bangalore, Karnataka, 560012, India.\\
  \textsuperscript{3}\,Institute for Solid State and Materials Research Dresden, Helmholtzstrasse 20, D-01069 Dresden, Germany\\
  \textsuperscript{4}\,Institut f\"ur Festk\"orperphysik,  Technische Universit\"at Dresden, D-01062 Dresden, Germany\\
  \textsuperscript{5}\,Department of Physics, University of California, San Diego, La Jolla, California 92093, USA.\\
   \textsuperscript{6}\,Abteilung Physikalische Chemie, Fritz-Haber-Institut der MPG, Faradayweg 4-6, D-14195 Berlin, Germany\\
   \textsuperscript{7}\,Max Planck Institute for Chemical Physics of Solids, D-01187 Dresden, Germany\\
   \textsuperscript{8}\,Institut f\"ur Physik, Universit\"at Augsburg, Universit\"atstr.1, D-86135 Augsburg, Germany\\
 \textsuperscript{9}\,Karlsruhe Institute of Technology, Institut f\"ur Festk\"orperphysik, D-76021 Karlsruhe, Germany.\\
 \textsuperscript{10}\,Van der Waals-Zeeman Institute, University of Amsterdam, NL-1018XE Amsterdam, the Netherlands\\
   }

\received{XXXX, revised XXXX, accepted XXXX} 
\published{XXXX} 


\abstract{%
%
%
%
\abstcol{In this article we review our angle- and time-resolved
photoemission studies (ARPES and trARPES) on various
ferropnictides. In the ARPES studies we focus first on the band
structure as a function of control parameters. We find a Lifshitz
transition near optimally "doped" compounds of hole/electron
pocket vanishing  type. Second we investigated the inelastic scattering rates
as a function of the control parameter. In contrast to the heavily
discussed quantum critical scenario we find no enhancement of the
scattering rate near optimally "doping". Correlation effects which
show up by the non-Fermi-liquid behavior of the scattering rates,
together with the Lifshitz transition offer a new explanation for
the strange normal state properties and suggests an interpolating
superconducting state between Bardeen-Cooper-Schrieffer (BCS)
and Bose-Einstein (BE) condensation.} {Adding
femtosecond time resolution to ARPES provides complementary
information on electron and lattice dynamics. We report on the
response of the chemical potential  to coherent optical
phonons in combination with
incoherent electron and phonon dynamics described by a three
temperature heat bath model. We quantify electron phonon coupling
in terms of $\lambda \langle\omega^2 \rangle$ and show that the
analysis of the electron excess energy relaxation is a robust
approach. The spin density wave ordering leads to  pronounced
momentum dependent relaxation dynamics. In the vicinity of $k_F$
hot electrons dissipate their energy by electron-phonon coupling
with a characteristic time constant of 200~fs. Electrons at the
center of the hole pocket exhibit a four times slower relaxation
which is explained by spin-dependent dynamics with its smaller
relaxation phase space. This finding has implications beyond the
material class of Fe-pnictides because it establishes experimental
access to spin-dependent dynamics in materials with spin density
waves.} }

%
%

\maketitle   

\section{Introduction}

The discovery of high-$T_c$ superconductivity in
LaO$_{1-x}$F$_x$FeAs \,\cite{Kamihara2008} has started a new era
in research on superconductivity \cite{Johnston2010,Kordyuk2012}.
In the following years several related families of iron-based
superconductors (FeSc\rq{}s) were discovered. They all have in
common that superconductivity is achieved by chemical substitution
or pressure starting from antiferromagnetic metallic parent
compounds, in which the important components are FeAs or FeSe/Te
layers. Since superconductivity appears at a quantum critical
point (QCP)\cite{Loehneysen2007,Gegenwart2008,Haslinger2002} at
the end of an antiferromagnetic or nematic region in a phase diagram
temperature vs. a control parameter, a wide spread opinion is that
superconductivity is mediated by antiferromagnetic  or nematic   quantum
fluctuations to which the charge carriers couple. Related to this
is the explanation for the observation of  non-Fermi-liquid (NFL)
behavior in the normal state above the QCP. Within this  scenario,
the linear temperature dependence of the resistivity observed in
the normal state of FeScs above the quantum critical point near optimal
doping can be related to a large coupling of the charge carriers
to the antiferromagnetic or nematic quantum
fluctuations\,\cite{Kasahara2010,Analytis2014}. Furthermore the
huge mass enhancement of the charge carriers detected in the
London penetration depth measurements near optimal doping was
explained by the existence of a quantum critical behavior in this
region of the phase diagram\,\cite{Analytis2014}. The
NFL behavior, which also appears in other
unconventional superconductors such as the heavy fermion systems
and the cuprates led  to the phenomenological theory of the
marginal Fermi liquid\,\cite{Varma2002} which describes the
many-body properties, i.e.,  the mass enhancement and the
scattering rate, of systems in which
the spectral weight of the coherent quasiparticles just disappears due to correlation effects.

\begin{figure}[htb]\centering
\includegraphics[width=\textwidth,angle=-90]{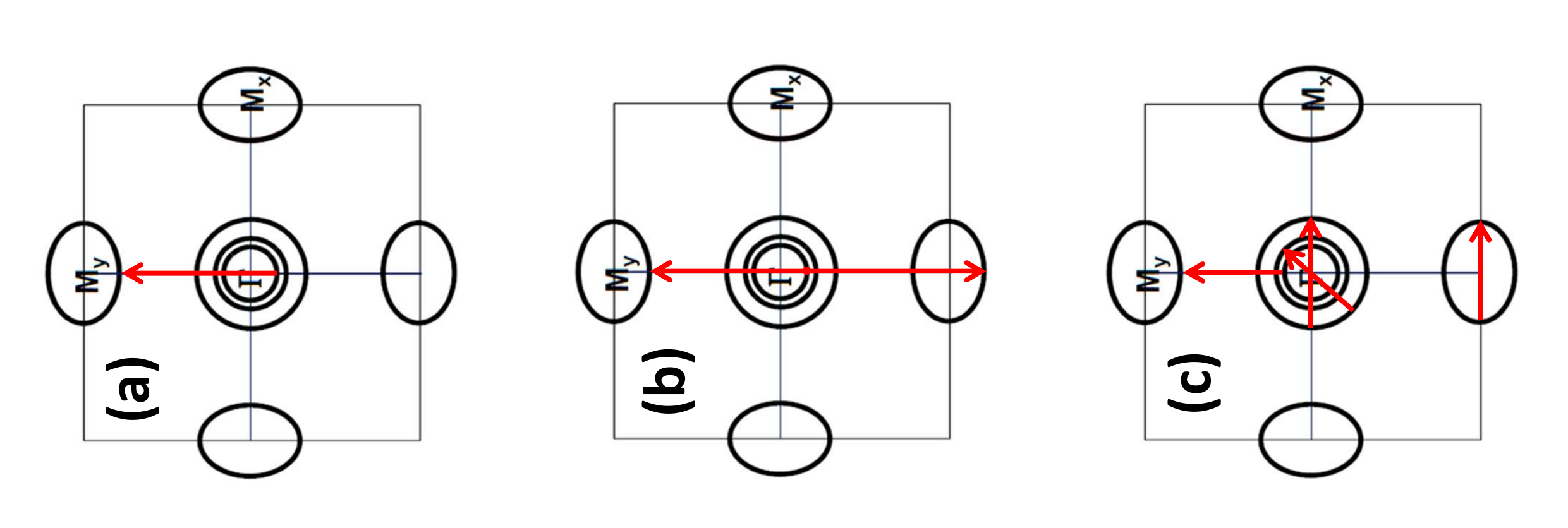}
\caption{ Schematic picture of the quasi two-dimensional Fermi
surface of ferropnictides. The red arrows present scattering
processes which lead to an antiferromagnetic or nematic susceptibility
(a), to a superconducting pairing susceptibility (b), and to
transport and thermal properties (c).} \label{scattering}
\end{figure}

Measurements of transport and thermal properties are valuable
methods to obtain information on the electronic structure of such
kind of solids. An excellent  overview on these properties  in the field
of ferropnictides was recently published
\,\cite{Rullier-Albenque2016}.  However, such measurements yield
information on the  properties  averaged over the entire  Fermi surface.
Even in the cuprates, where there is only one band close to the
Fermi surface, momentum dependent information is needed since the
many-body properties change when moving around the Fermi surface.
In the case of the FeScs, there are at least
four Fe3$d$ bands close to the Fermi surface which form hole pockets  in the center and
electron pockets at the edges of the Brillouin zone (BZ) in this
quasi two-dimensional systems. In
Fig.\,\ref{scattering} we present a schematic drawing of the Fermi
surface together with scattering processes.

\begin{figure}[t]\centering
\includegraphics[width=\columnwidth]{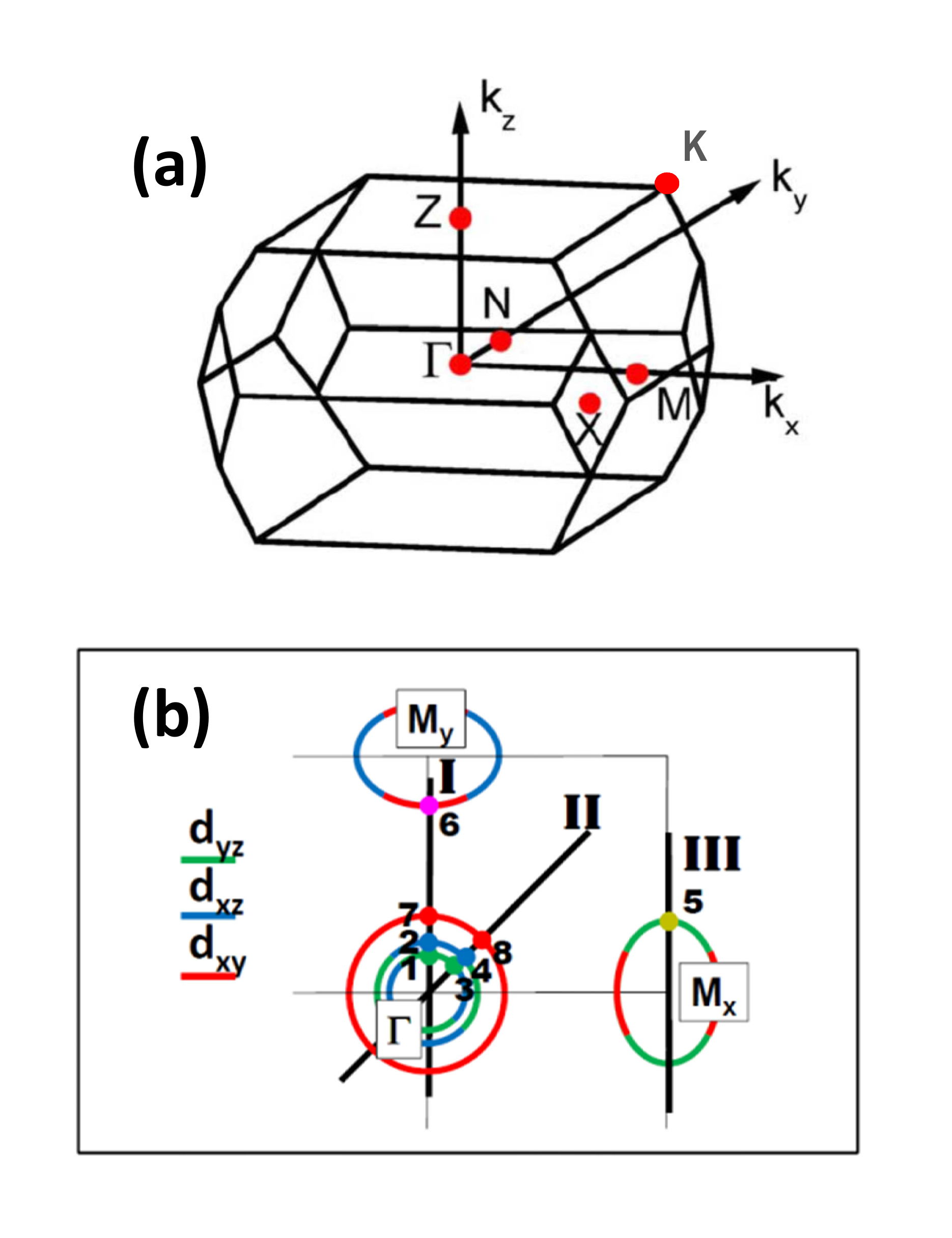}
\caption{  (a) Brillouin zone of  ferropnictides. (b)  Schematic
Fermi surface of ferropnictides in the $k_z=0$ plane together with
cuts  performed in the APES and trARPES experiments. The orbital
character of the Fermi surfaces was adopted from \cite{Kemper2011}
and is marked  in different sections  by different colors. Points
at which scattering rates were determined are labeled by numbers
and different colors.} \label{Brillouin}
\end{figure}

Since there are very often very strong nesting conditions between
hole and electron pockets, the antiferromagnetic and the pairing
susceptibility are believed to be dominated by interpocket scattering processes
between the hole pockets and the electron pockets. This led to the
proposal of  s+- superconductivity, in which the superconducting
order parameter changes the sign when moving from the hole pockets
to the electron pockets\,\cite{Mazin2008a}. It is expected that
the antiferromagnetic order and the superconducting order are
mediated by strong inter- and intra-pocket scattering processes (between hot spots) while
the conductivity is related to weak scattering processes (between
cold spots). In order  to obtain a microscopic understanding of
superconductivity and the normal state properties (e.g. the
conductivity), it is important to obtain information of the
scattering  process between the various parts of the  Fermi surface and even
between different sections of the Fermi surfaces. Angle-resolved
photoemission spectroscopy (ARPES)\,\cite{Damascelli2003} and
time-resolved angle-resolved photoemission spectroscopy (trARPES)
are especially valuable methods since they provide
momentum-dependent information in the multi-band FeScs
for the occupied and unoccupied states, respectively, in the relevant
energy range around the Fermi level. In the
present review we present measurements near the high symmetry
points in various FeScs. This is illustrated in
Fig.\,\ref{Brillouin},  where we show the three-dimensional BZ of
ferropnictides together with a cut of the band structure at the
Fermi level at $k_z=0$. In the ARPES experiments we have performed
three cuts (I, II, and III) to reach the points 1-8. In the
trARPES experiments we have measured along cut I in the region of
the hole pockets.

In the present review we report on results from the compounds
\BFCA , \BFAP , \EFAP , \NFCA , and \NFRA . These compounds are
usually termed 122 and 111 compounds.

\begin{figure*}[t]%
 \centering
\includegraphics*[width=\textwidth,angle=0]{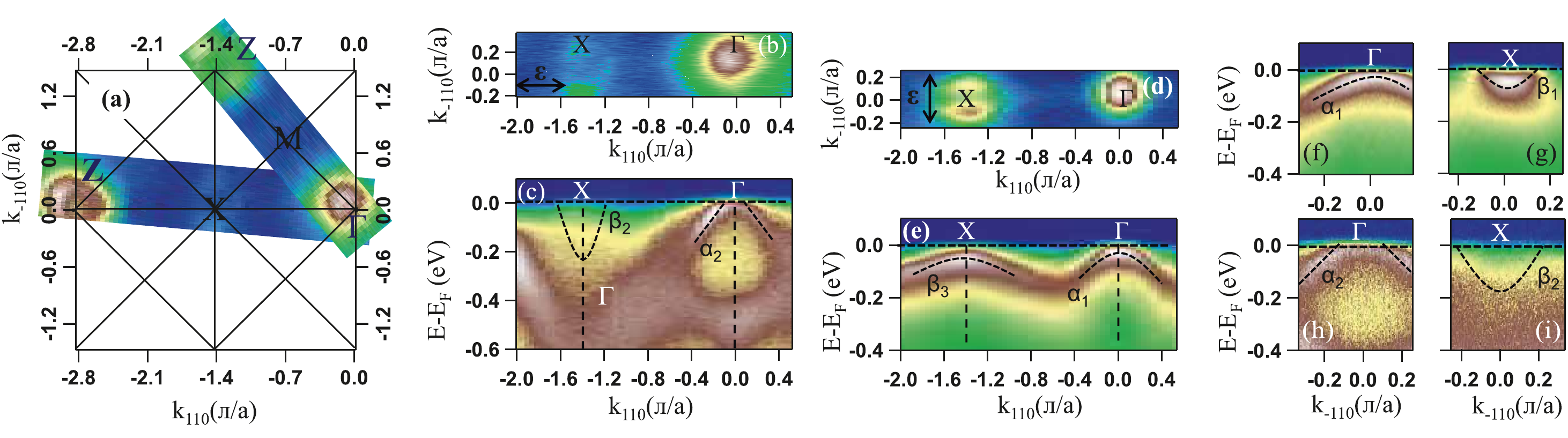}
\caption{ARPES data of a slightly underdoped \BFCA\ $x=0.04$
single crystal. (a) Fermi surface map taken at the high symmetry
points $\Gamma$, $M$,
$X$ and $Z$ projected on to a tetragonal two dimensional BZ. (b)
Fermi surface taken from (a) along $k_{110}$ direction passing
through the  $X$ and $\Gamma$ points. (c) Energy distribution map
(EDM) taken from (b) in order to reveal the band dispersion along
the $\Gamma$-$X$ high symmetry line . (d) and (e) Fermi surface
map and EDM through the $X$ and $\Gamma$-points, respectively,
measured using vertical polarization . (f) and (h) EDMs taken at
$\Gamma$ along the $ k_{−110}$ direction measured using vertical
and horizontal polarization, respectively. (g) and (i)  EDMs taken
at $X$ along the $ k_{−110}$ direction measured using vertical
and horizontal polarization, respectively. Reprint
from\,\cite{Thirupathaiah2011a}.} \label{cobalt}
\end{figure*}

\section{Experimental details}

\subsection{Single crystals}

Using the self-flux method single crystals of \BFCA \, were grown
in Amsterdam. Characterization of these samples was reported in
\cite{Massee2009}. Another set of \BFCMA \, single crystals was
grown at the KIT from the self-flux and characterized by
resistivity, magnetization, specific heat, and dilatometry
measurements\,\cite{Hardy2009}. Single crystals of \BFAP\  and
\EFAP\ were grown in G\"ottingen and Augsburg using the Sn-flux and
the Bridgman (without Sn flux) method, respectively and they were
characterized by various methods \cite{Jeevan2011}. Single
crystals of \NFCA\ , \NFRA\ , and \BCA\ were grown in Dresden using as well
the self-flux method \cite{Steckel2015}.

\begin{figure*}[t]%
\includegraphics[width=\textwidth,angle=0]{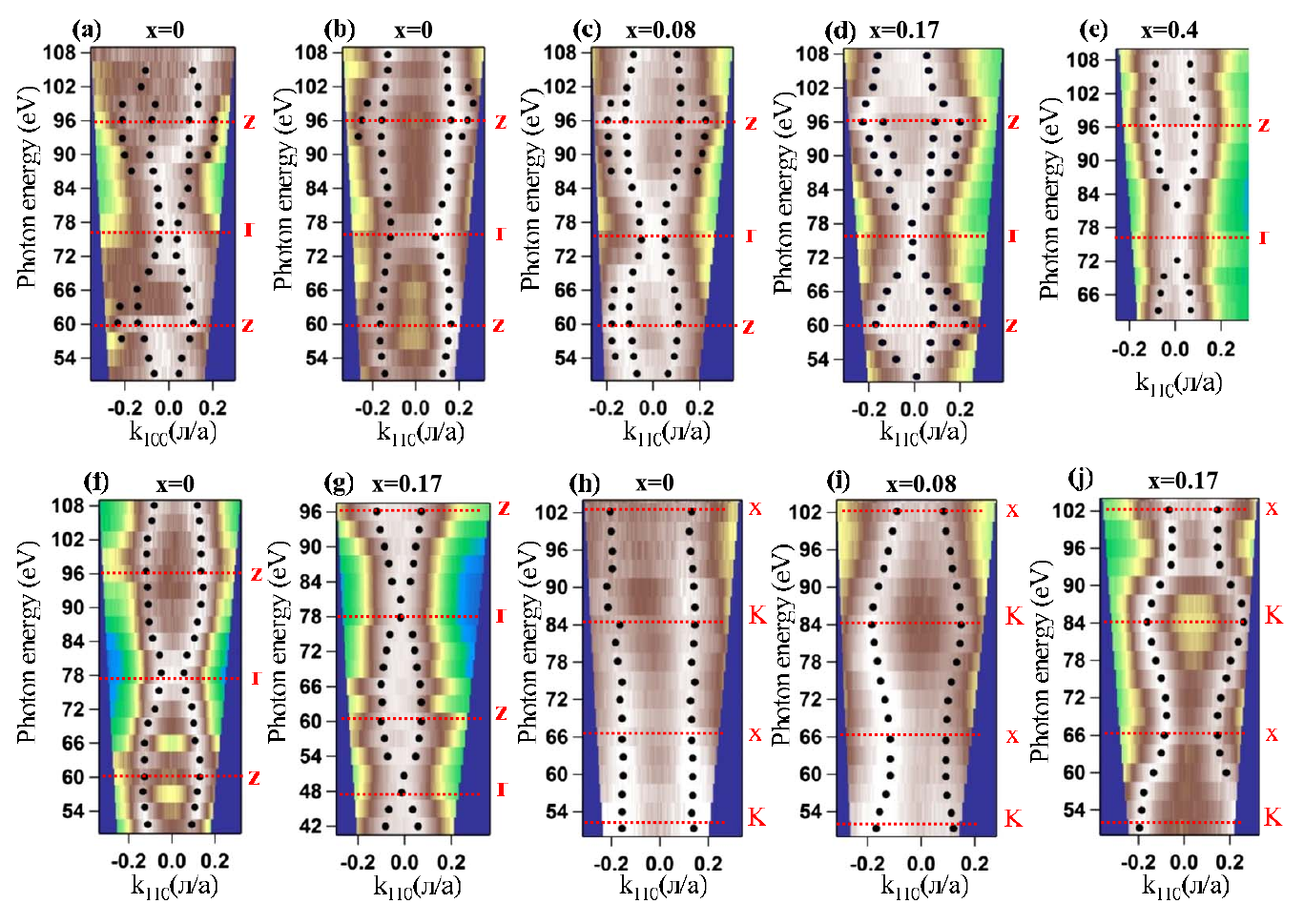}
\caption{ Photon energy dependent ARPES measurements performed on
\BFCA\ to reveal the $k_z$ dispersion as a function of doping
concentration. (a) Fermi surface for $x=0$, in the $k_{100}$ vs
h$\nu$ plane near $ k_x=k_y=0$ measured with p-polarized photons.
(b) Analogous data as in (a)  but in the $k_{110}$ vs h$\nu$
plane. (c), (d), and (e) Analogous data as in (b) but for x=0.8,
0.17, and 0.4, respectively. (f) and (g) Analogous data as in (b)
but measured with s-polarized photons for $x=0$ and 0.17. (h),
(i), and (j) Fermi surfaces in the $k_{110}$ vs h$\nu$  plane near
$k_x=k_y=1$ measured with s-polarized photons for $x=0, 0.08$, and
$0.17$, respectively. Reprint from\,\cite{Thirupathaiah2010}.
Copyright 2010, American Physical Society.} \label{cokz}
\end{figure*}

\subsection{ARPES}

ARPES measurements were conducted at the $1^2$  and $1^3$  ARPES
endstations attached to the beamline UE112 PGM 2 at BESSY with
energy and angle resolutions between 4 and 10 meV and $0.2
^\circ$, respectively. Variable photon energies h$\nu = 20 - 130$
eV were used to reach different $k_z$ values in the BZ. The use of
polarized photons provides information on the orbital character of
the electronic states by matrix element effects\,\cite{Fink2009}.
If not otherwise stated the measuring temperature at the $1^2$ and
$1^3$ ARPES endstations were 30 and 0.9 K, respectively.
Temperatures above the N\'eel temperature were used for
unsubstituted or weakly doped  samples in order to keep the
compounds in the paramagnetic state and and to avoid back-folding
of the bands due to the antiferromagnetic order\,\cite{Jong2010}.

Although ARPES is a highly surface sensitive method, previous
experimental and theoretical studies using density functional
theory (DFT) band structure calculations  show that ARPES results
are close to the bulk electronic structure. In the 122 compounds a
disordered Ba/Eu surface layer is formed upon cleaving at low
temperatures  which leads to additional broadening of the lines
due to elastic scattering\,\cite{Heumen2011}. Ordering at higher
temperatures leads to backfolding of  bands. In the 111 compounds,
calculations indicate that the electronic structure of the surface
layer is close to that of the  bulk\,\cite{Lankau2010}.

\subsection{trARPES}
Femtosecond time- and angle-resolved ARPES measurements reported
here were carried out in the Physics Department of the FU Berlin
using the experimental setup described in
\cite{Lisowski2004,Schmitt2011}. Employing a 300~kHz commercial
Ti:sapphire amplifier (Coherent RegA 9050) femtosecond laser
pulses at 1.5~eV fundamental photon energy and 55~fs pulse
duration were generated. These pulses are used as pump pulses in
pump-probe experiments. The probe pulses with photon energy of
6.0~eV are generated by splitting off a fraction of the
fundamental pulse which is subsequently frequency quadrupled in
two subsequent BBO non-linear optical crystals, remaining
synchronized with the pump pulses. The obtained sub 100~fs probe
pulse duration determines the overall time resolution of the
experiment. Temporal and spatial overlap of pump and probe pulses
are obtained in ultrahigh vacuum on the sample surface of in-situ
cleaved Fe-pnictide single crystals. The energy resolution of the
experiment of 50~meV is set by the laser pulse bandwidth required
to generate fs laser pulses. To detect the low energy
photoelectrons of about 1~eV kinetic energy an electron
time-of-flight spectrometer is used in combination with single
electron counting. For sample cooling to 25~K a lHe cryostat was
employed.

\begin{figure}[t]%
\centering
\includegraphics[width=\columnwidth]{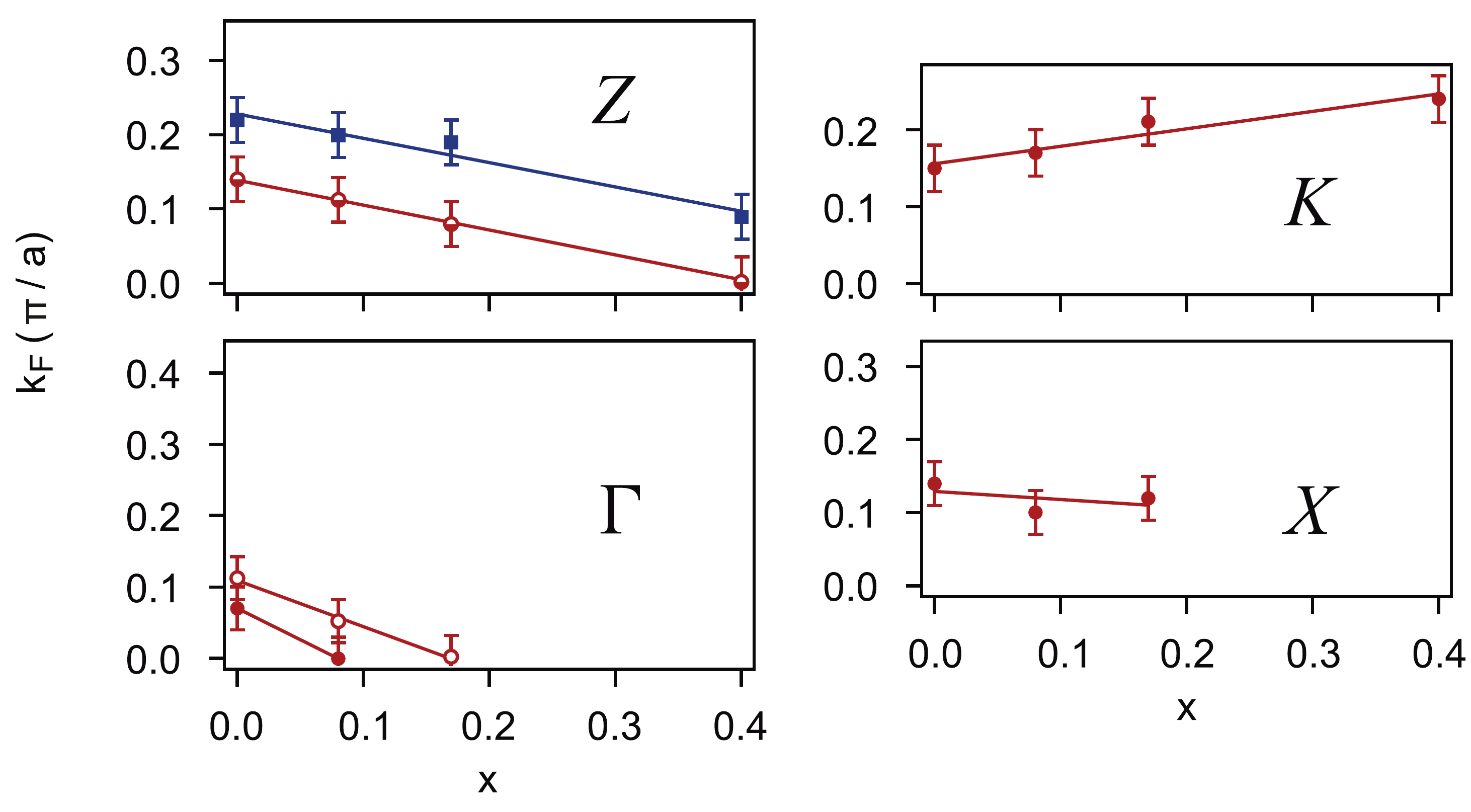}
\caption{Fermi vectors of \BFCA\ as a function of Co
doping near the high symmetry points $Z, \Gamma ,K ,$ and $X$. The
orbital character of different bands is marked by different
symbols: open (closed) circles predominantly  $xy (xz/yz)$
character, closed squares predominantly $z^2$ character. Reproduced
from\,\cite{Thirupathaiah2011} }
\label{kF}
\end{figure}

\section{ARPES results}

\subsection{Band structure, Fermiology, Lifshitz transitions}

In Fig.\,\ref{cobalt} we show representative ARPES data of a
slightly underdoped \BFCA\ $x=0.04$ single
crystal\,\cite{Thirupathaiah2011,Thirupathaiah2010}.
Fermi surfaces were derived from an integration of the spectral
weight close to the Fermi level. In the energy distribution maps
(EDMs) we see -- depending on the polarization --  the inner hole
pocket ($\alpha_1$) or the middle hole pocket ($\alpha_2$)
near $\Gamma$.
The intensity of the outer hole pocket with predominantly $3d xy$ orbital
character is very weak in this compound because bands with this
orbital character exhibit the highest elastic scattering
rates\,\cite{Herbig2015}. In addition,    electron pockets
($\beta_1$ and $\beta_2$) are detected near $X$.

In Fig.\,\ref{cokz} we present photon energy dependent data
measured near the $\Gamma$ and the $X$  point which provide
information on the $k_z$ dispersion of the bands
\,\cite{Thirupathaiah2010}. In agreement  with density functional
theory (DFT) band structure calculation \cite{Thirupathaiah2010},
a modest  doping dependent $k_z$ dispersion is detected in the
ARPES data for the hole pockets and a very weak one for the
electron pockets.

\begin{figure}[t]%
\includegraphics[width=\columnwidth]{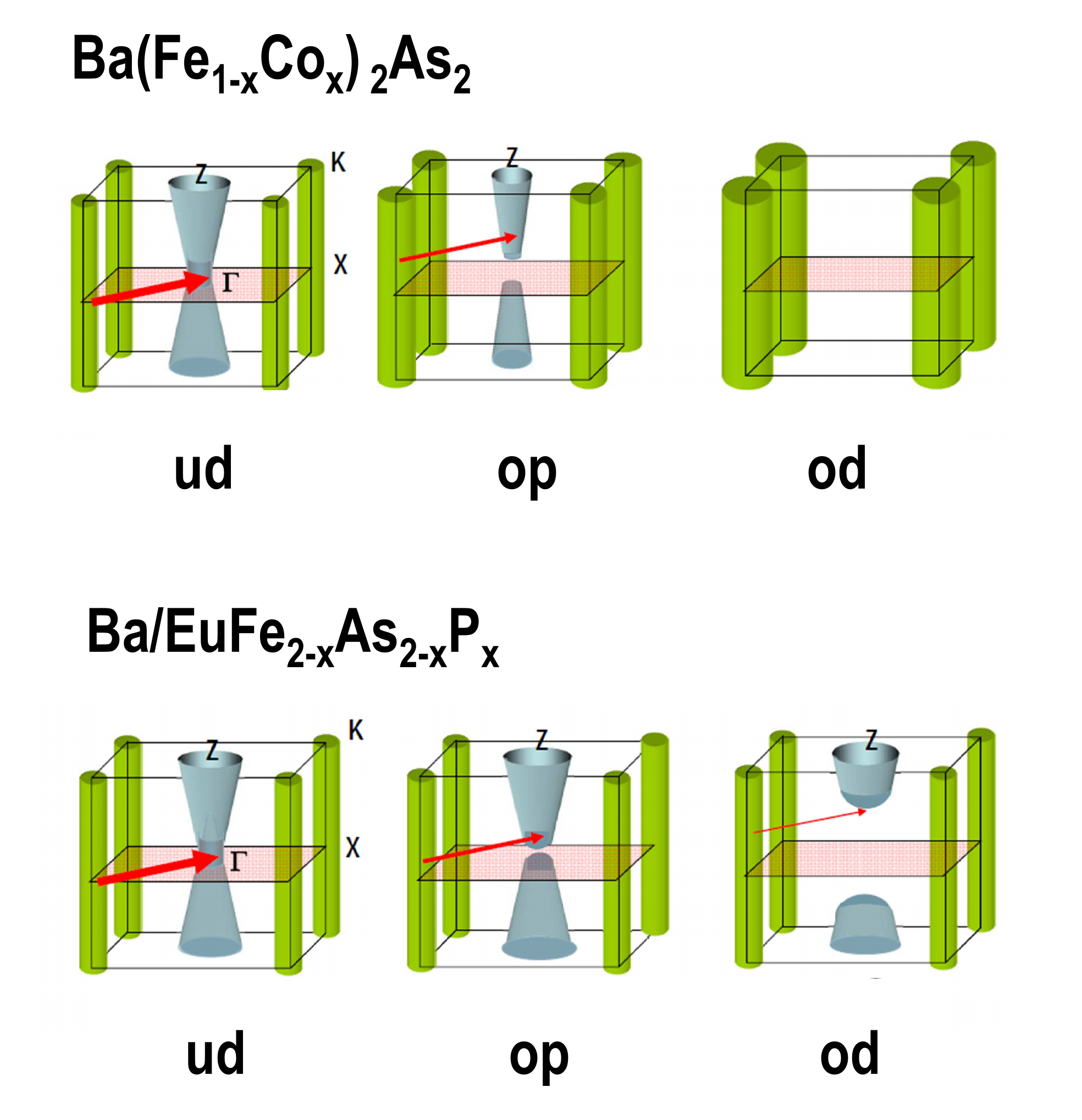}
\caption{Schematic illustration of the Fermi cylinders of the
$xz/yz$ hole (blue) and electron pockets (green) in \BFA\  as
function of electron doping  by Co substitution of Fe (upper
panel) and chemically pressurizing by isovalent P substitution of
As (lower panel). ud: undoped, op: optimaly doped/substituted, od:
overdoped/oversubstituted.} \label{Fermicyl}
\end{figure}

It is interesting to look closer at the doping dependent band
structure. As presented in Fig.\,\ref{cokz} the size of the Fermi
surface in the $k_x-k_y$ plane at  $\Gamma$ decreases rapidly with
increasing  Co concentration while the reduction at the $Z$  point
is weaker. This result is also obtained in a more refined
evaluation of the $k_F$ values shown in Fig.\,\ref{kF}
\,\cite{Thirupathaiah2010}.  In particular at the  $\Gamma$ point
the $k_F$ values  shown in Fig.\,\ref{cobalt} show  a strong
concentration dependence. There the top of the inner hole pocket
moves through the Fermi level near optimal doping ($x=0.06$) ,
i.e., where the Fermi wave vector goes to zero (Fig.\,\ref{kF}
panel $\Gamma$) and  where the Fermi surface disappears
(Fig.\,\ref{cokz} (d) and (g)). Thus for the inner hole pocket
there is a Lifshitz transition of the pocket vanishing
type\,\cite{Lifshitz1960} at the $\Gamma$ point: the Fermi
cylinder transforms into an ellipsoid around the $Z$ point (see
Fig.\,\ref{Fermicyl}).  As a consequence of the Lifshitz
transition, there are flat bands at the Fermi level.  In the
overdoped region,  the length of the ellipsoid along the $k_z$
line is getting shorter and shorter and finally at the end of the
superconducting region  the inner hole pocket becomes completely filled
also at the $Z$  point (see Fig.\,\ref{Fermicyl}).

\begin{figure}[t]%
\centering
\includegraphics[width=0.7\columnwidth,angle=0]{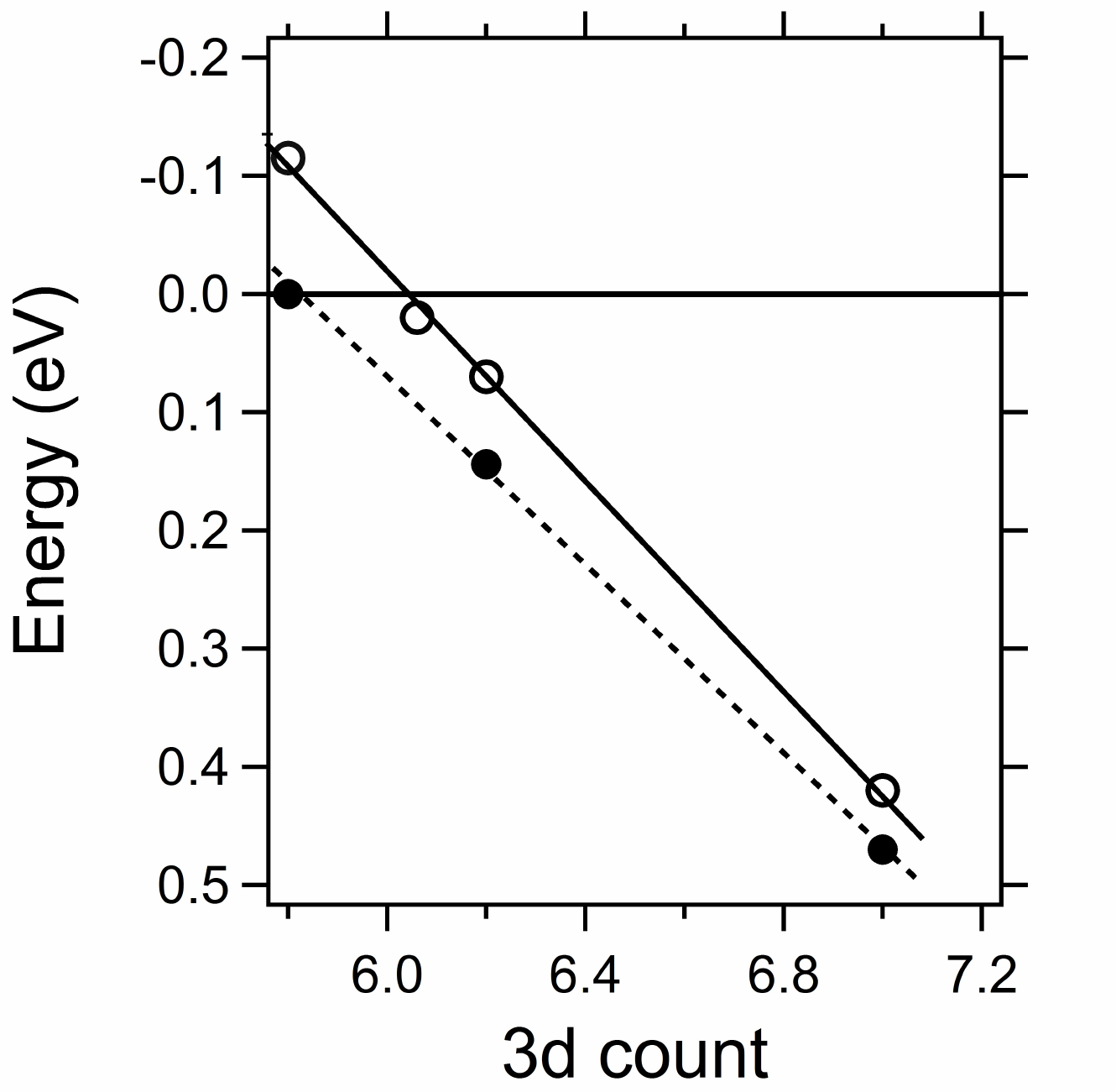}
\caption{The top of the inner hole pocket in \BFCA\ compounds
(open circles) and the bottom of the electron pocket (closed
circles) as a function of 3$d$ count. Data from nearly optimally
hole doped  \KBFA\  are added. } \label{E0vs3d}
\end{figure}

In Fig.\,\ref{E0vs3d} we plot the top of the inner hole hole
pocket and the bottom of the electron pocket having predominantly
both $yz$ character at the points 1 and 5 in
Fig.\,\ref{Brillouin}(b) as a function of $3d$ count, i.e., the
number of $3d$ electrons in the transition metal ion. We focus to
these points  since at these points the largest superconducting
gaps have been detected together with the largest scattering rates
(see below). For the electron doped ferropnictides the top of the
inner hole pocket moves through the Fermi level very close to the
optimal doping concentration, while for the hole doped systems the
bottom of the electron pocket moves through the Fermi level at
optimal doping. Thus we have a Lifshitz transition
of bands having $yz$ character in both cases.

Very similar data were derived for the isovalently substituted
compounds \BFAP\ and \EFAP\ \cite{Thirupathaiah2011}. Also in these
systems a Lifshitz transition of the inner hole pocket is observed
near optimal substitution. Different from  the electron doped
compound is that in order to achieve charge neutrality, the Fermi
surface at the $Z$ point increases  (see Fig.\,\ref{Fermicyl}).
Thus even at the highest P concentration there is still a partially empty hole
pocket having $xz/yz$ character.  Possibly this observation is
related to the fact that superconductivity is observed even in the
pure \BFP\ compound.

\subsection{Scattering rates}

\begin{figure}[t]
\centering
\includegraphics[width=\linewidth,angle=0]{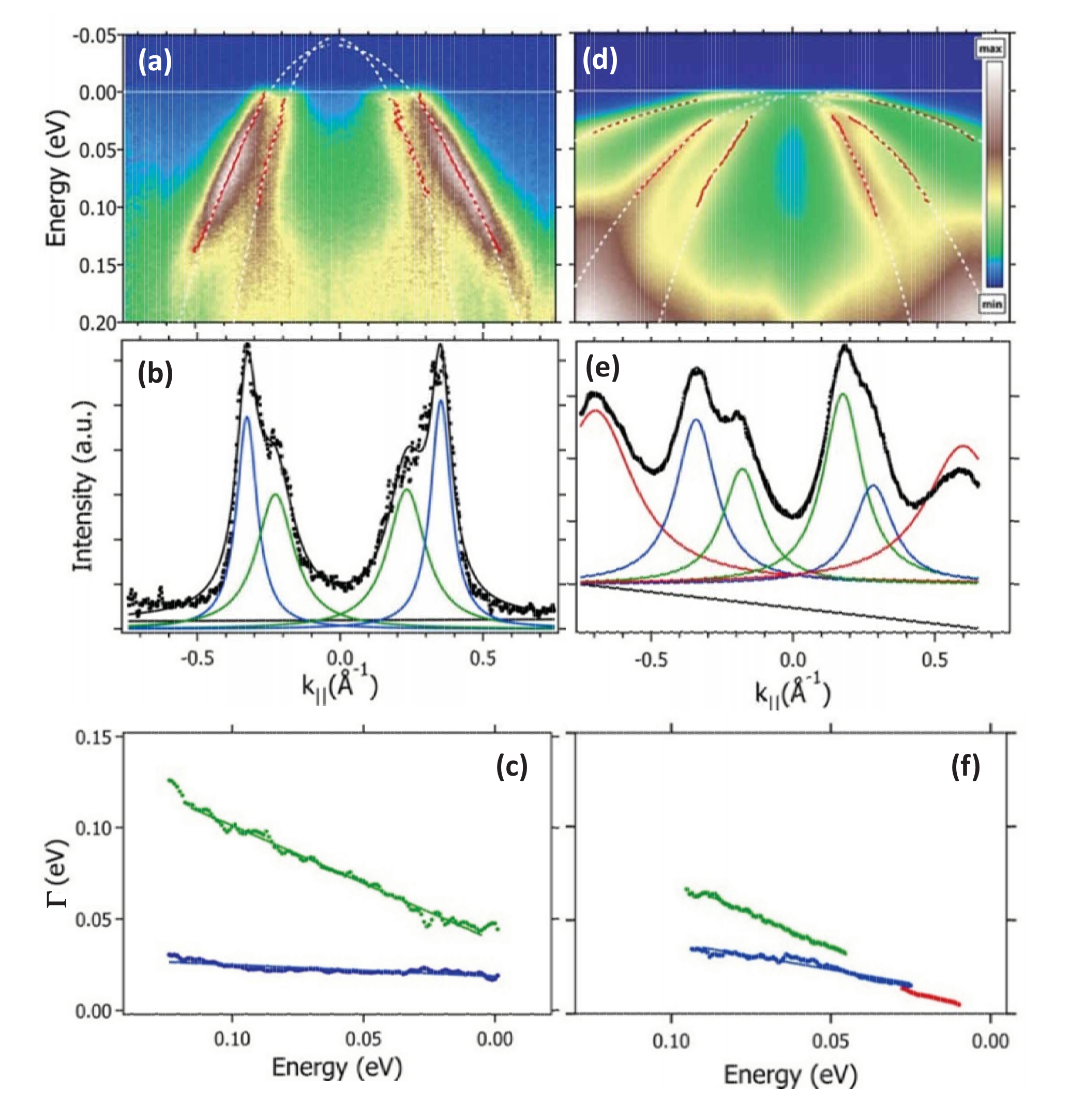}
\caption{(a) and  (d) ARPES energy distribution maps of \BFAP\
$x=0.27$  and \NFRA\ $x=0.027$ , respectively, measured along cut
I. The red lines depict dispersions derived from the momentum
distribution curves, which are shown in (b) and (e) for a binding
energy of 40 meV. (c) and (f) Experimental  scattering rates
$\Gamma (E)$ for  \BFAP\  $x=0.27$  and \NFRA\ $x=0.027$,
respectively, together with a linear fit. Green: near point 1,
blue: near point 2 , red: near point 7 (see
Fig.\,\ref{Brillouin}(b))\,\cite{Fink2015}.
 }
 \label{MDCs}
\end{figure}

In this section we review our ARPES results on the scattering
rates in ferropnictides. In Fig.\,\ref{MDCs}(a) and (d) we show
typical energy distribution maps (MDCs) of optimally
doped/substituted \BFAP\ and \NFRA\ . In (a) the inner two hole
pockets and in (d) all three hole pockets can be clearly resolved.
Typical momentum cuts (MDCs) together with fits by Lorentzians are
shown in  Fig.\,\ref{MDCs}(b) and (e). In Fig.\,\ref{MDCs}(c) and
(f) we present the scattering rates  $\Gamma (E)$ derived from the
width in momentum space\,\cite{Fink2015}. For the hole pockets in
\BFAP\ we find in a large energy range of $5 \le E \le 120$  meV a
NFL linear-in-energy dependence, which can be described by
$\Gamma (E)=\alpha+\beta E$. Due to a finite energy and momentum
resolution 5 meV is the lower boarder of the range in which
reasonably values for the scattering rates can be derived in the
present ARPES experiments. Similar results were obtained for
\BFCA\ and \BFCMA\  \cite{Rienks2013}. Since in \NFRA\ there is a
strong hybridization between bands close to the Fermi level, the
energy ranges for which the scattering rates can be determined, is
restricted.  $\alpha$ is a measure for elastic scattering
processes, e.g. scattering by impurities. $\beta$  is a measure of
the strength of inelastic scattering processes. In the inelastic
scattering rates or the imaginary part of the self-energy no
evidence for a strong coupling to bosonic excitations such as
phonons is detected. If phonons would determine the scattering
rate  a step like increase should be observed at the phonon
energies (which are smaller than 40 meV)\,\cite{Engelsberg1963}.
Moreover, in the present ARPES data no kinks in the  dispersion are realized
which are expected when the charge carriers are strongly coupled
to phonons\,\cite{Damascelli2003}. Thus the observed energy dependence clearly shows that
the scattering rates are predominantly determined by electronic excitations, i.e.,
by an Auger process in the
valence band: a relaxation of the photo-electron hole to lower binding
energies  and a transfer of the received energy to an
electron-hole excitation\,\cite{Mahan2000}.

Differences in the inelastic scattering rates have been predicted
by theoretical calculations\,\cite{Graser2009,Kemper2011}:
scattering rates between sections of the Fermi surface having the
same orbital character (e.g. from point 1 to point 5 in
Fig.\,\ref{Brillouin}(b) having both predominantly $yz$ orbital
character) should be bigger than scattering rates between sections
having a non-equal orbital character (e.g. from point 2 to point 5
in Fig.\,\ref{Brillouin}(b) having predominantly $xz$ and $yz$
orbital character, respectively). The reason for this is that the
former is determined by the on-site Coulomb repulsion $U$, while
the latter is reduced to  $U-J$ due to the Hund\rq{}s exchange
coupling $J$  which appears when considering transitions between
parallel spin states. Parallel spins should occur in the case when
the orbital character of the initial and the final state in the
scattering process are antisymmetric.

The linear dependence of the scattering rates, i.e.,
NFL behavior, indicate a considerable amount of
correlation effects in the Fe$3d$ system. Such a behavior is often
described by the phenomenological model of a  marginal Fermi
liquid\,\cite{Varma2002}.

\begin{figure}[t]%
\centering
\includegraphics[width=\columnwidth,angle=-90]{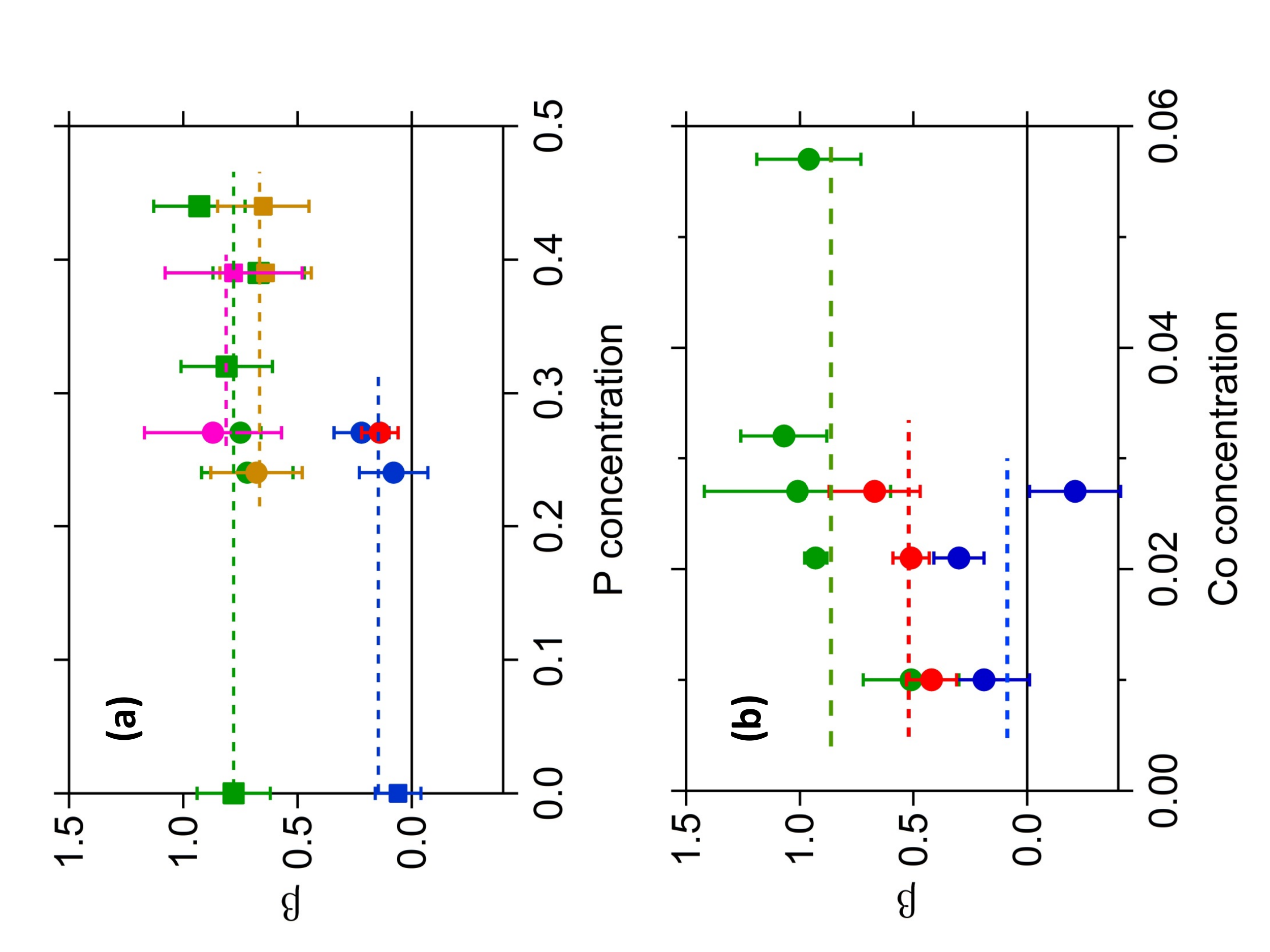}
\caption{Doping/substitution dependent $\beta$ values  in \BFAP\
and \EFAP\ (upper panel), and  \NFCA\ and \NFRA\ ( lower panel)
for various high-symmetry points near $\Gamma$ and $X$. The color
code of Fig.\,\ref{Brillouin}(b) for the high-symmetry points is
used. Reproduced from\,\cite{Fink2015} }\label{beta}
\end{figure}

In Fig.\,\ref{beta} a compilation of all available  $\beta$ values
for hole and electron pockets in \BFAP\ and \EFAP\ , in \NFCA\ and
\NFRA\ are presented. Interestingly no enhancement of the
scattering rate is observed near optimal doping. This is in stark
contrast to what is naively expected in the quantum critical
scenario:  the scattering rates should be strongly enhanced near
the QCP due to a coupling to antiferromagnetic or nematic  quantum
fluctuations.

\begin{figure}[t]%
\centering
\vspace{-1cm}
\includegraphics[width=1.1\linewidth,angle=0]{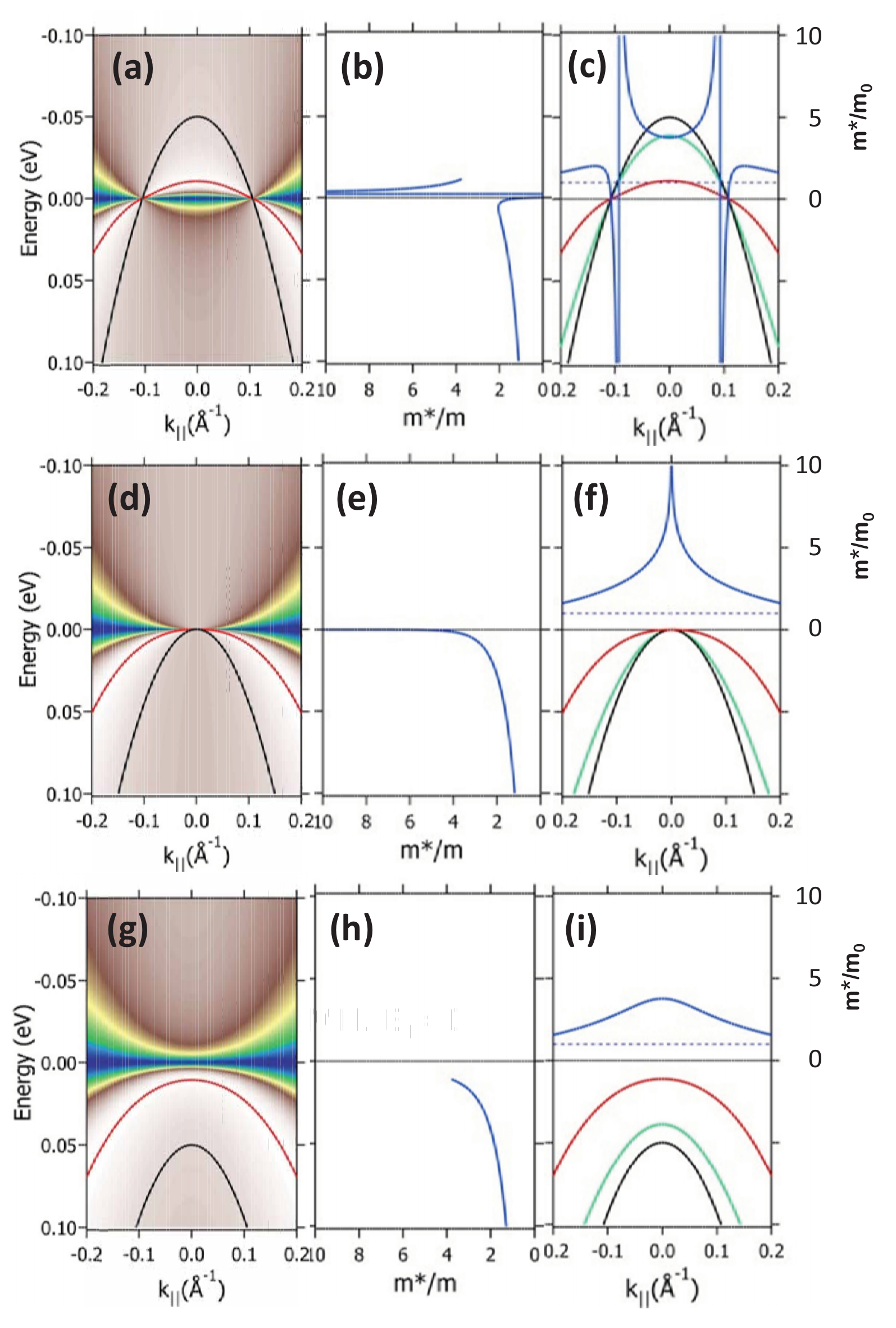}
\vspace{-1cm} \caption{Calculation of electronic structure data
for FeScs. Left column: false color map of the spectral function
together with the bare particle dispersion (black) and the
renormalized dispersion (red). Middle column: energy dependent
effective mass $m^*/m_0$ (blue). Right column: bare particle
dispersion (black), $\Re\Sigma$ (green) and renormalized
dispersion (red). Top row: $E_t=-50$ meV, simulating underdoped
compounds; middle row $E_t=0$, simulating optimally doped
compounds close to a Lifshitz transition; bottom row: $E_t=-50$
meV, simulating overdoped compounds. } \label{mass}
\end{figure}

\subsection{Coaction of Lifshitz transitions and correlation effects.}

Recently an explanation of the appearance of a linear-in-energy
increase of the scattering rate was presented pointing out the
importance of a Lifshitz transition\,\cite{Fink2016}. A simple
calculation of the $\Im\Sigma$, in which a constant density of
states at the Fermi level was assumed, as a function of the
extension  of the unoccupied part of the band above $E_F$  yielded
a NFL behavior near the Lifshitz transition, i.e., when the the
top of the hole  pocket ($E_t$) is close to the Fermi level.
Decreasing  $E_t$, i.e.,  increasing the energy difference
$E_F-E_t$ and thus moving away from the Lifshitz transition,
extends the range of Fermi liquid behavior to higher energies. On
the other hand, a decrease of an $|E_F-E_t|$ or an increase of
correlation effects which renormalizes the width of unoccupied
states and thus reduces also $|E_F-E_t|$, moves the NFL behavior
to lower energies. If $E_t$ is related to a control parameter and
correlation effects are important , this means that NFL behavior
should appear at lower energies and in a wider range of the
control parameter. This is a consequence of a constant
susceptibility for electron-hole excitation which follows from a
convolution of a delta-function like unoccupied density of states
with a constant occupied density of states.

This idea was supported by further evaluation of our ARPES
data. As shown above, from these data we could derive $\Im\Sigma$.
Since this function is connected to $\Re\Sigma$ by the
Kramers-Kronig relation, we could calculate the latter from our
ARPES data. In these calculations the experimental $\Im\Sigma$ was
extrapolated linearly to zero energy and at high energy a cutoff
energy of $E_c=1.5$ eV was introduced. Alternatively, we used the
parameters which describe $\Im\Sigma$ and use the marginal Fermi
liquid model to calculate $\Re\Sigma$. In this model the
self-energy is given by\,\cite{Varma2002}

\begin{equation}
\Sigma=\frac{1}{2}\lambda_{MF}E ln\frac{E_c}{E}-i\frac{\pi}{2}\lambda_{MF}E.
\end{equation}

In order to obtain representative results for the FeScs we used in
the calculations presented in Fig.\,\ref{mass} a coupling constant
of $\lambda_{MF}=1.4$. This value is between those derived for the
hole doped and the electron doped compounds\,\cite{Fink2016a}. The
renormalized dispersion is derived by a subtraction of   the
calculated $\Re\Sigma$ from the bare particle dispersion derived
from DFT band structure calculations. In Fig.\,\ref{mass} we
present such calculations as a function of three values of
$E_F-E_t$, simulating  underdoped (upper row), optimally doped
(middle row), and overdoped (lower row) compounds. In the left
column a calculation of the spectral function is presented. In the
middle column the derived effective mass $m^*/m_0$ as a function
of the energy is depicted. In the right column we show the bare
particle dispersion (black),  the $\Re\Sigma$ (green), and the
renormalized dispersion (red). In addition we show in the right
panels the effective mass $m^*/m_0$ by a blue line,  as a function
of momentum. For $E_t=E_F$, where the top of the bare particle
dispersion just touches the Fermi level, i.e.,  when  at the Fermi
level the dispersion of the bare particle band is flat (see
Fig.\,\ref{mass}(f)), the $\Re\Sigma$ is very close to the bare
particle band and therefore the renormalized dispersion at the
Fermi level is strongly reduced. This corresponds to a very high
effective mass $m^*/m_0 \approx 8$ at $k=0$ (see
Fig.\,\ref{mass}(f)). In Fig.\,\ref{mass}(e)),  a similar mass
enhancement appears close to the Fermi level. The  spectral
function (Fig.\,\ref{mass}(d)) is dominated by incoherent spectral
weight, i.e., by  particles for which
the life-time broadening is larger than the binding energy.
Interestingly, a large completely incoherent mirror like spectral
weight is observed as a satellite structure above the Fermi level.
Together with a weak quasiparticle structure below the Fermi
level, the incoherent spectral weight below and above $E_F$ allows
strong intra pocket and inter pocket scattering with electron
pockets which leads to large low-energy scattering rates which are
characteristic for marginal Fermi liquids and which leads to the
strong mass enhancement at low energies. Moving away from $k=0$ or
$E=E_F$, the effective mass is strongly reduced to a value of
$m^*/m_0 \approx 2$.

Changing $E_t$ to -50 meV or 50 meV, corresponding to a change of
the control parameter to underdoped and overdoped compounds,
respectively, strongly reduces the size of the effective mass (see
Fig.\,\ref{mass}(a), (b), (e), and (f)). As expected the
incoherent spectral weight far above the Fermi level is not
changing very much as a function of the control parameter.

A combination of the band structure results and the calculation of
the effective mass on the basis of experimental scattering rates
leads to the following explanation of the NFL behavior revealed in
the ARPES experiments. Using the $3d$ count dependence of $E_t$
leads in \BFCA\ to a concentration dependence  $E_t=0.4x$. This
would lead to a NFL behavior above our resolution determined low
energy boarder of 5 meV above the energy $ |0.4(x-0.06)|$ which
would lead to a NFL behavior in a narrow concentration range of
$x=0.06 \pm 0.0125$. On the other hand, a low energy mass
enhancement of eight would result in a NFL behavior in a much
larger concentration range of about $x=0.06 \pm 0.1$. This
explains our ARPES results of a NFL behavior above 5 meV in a
large concentration range.

\begin{figure}[t]%
\includegraphics[width=\linewidth,angle=0]{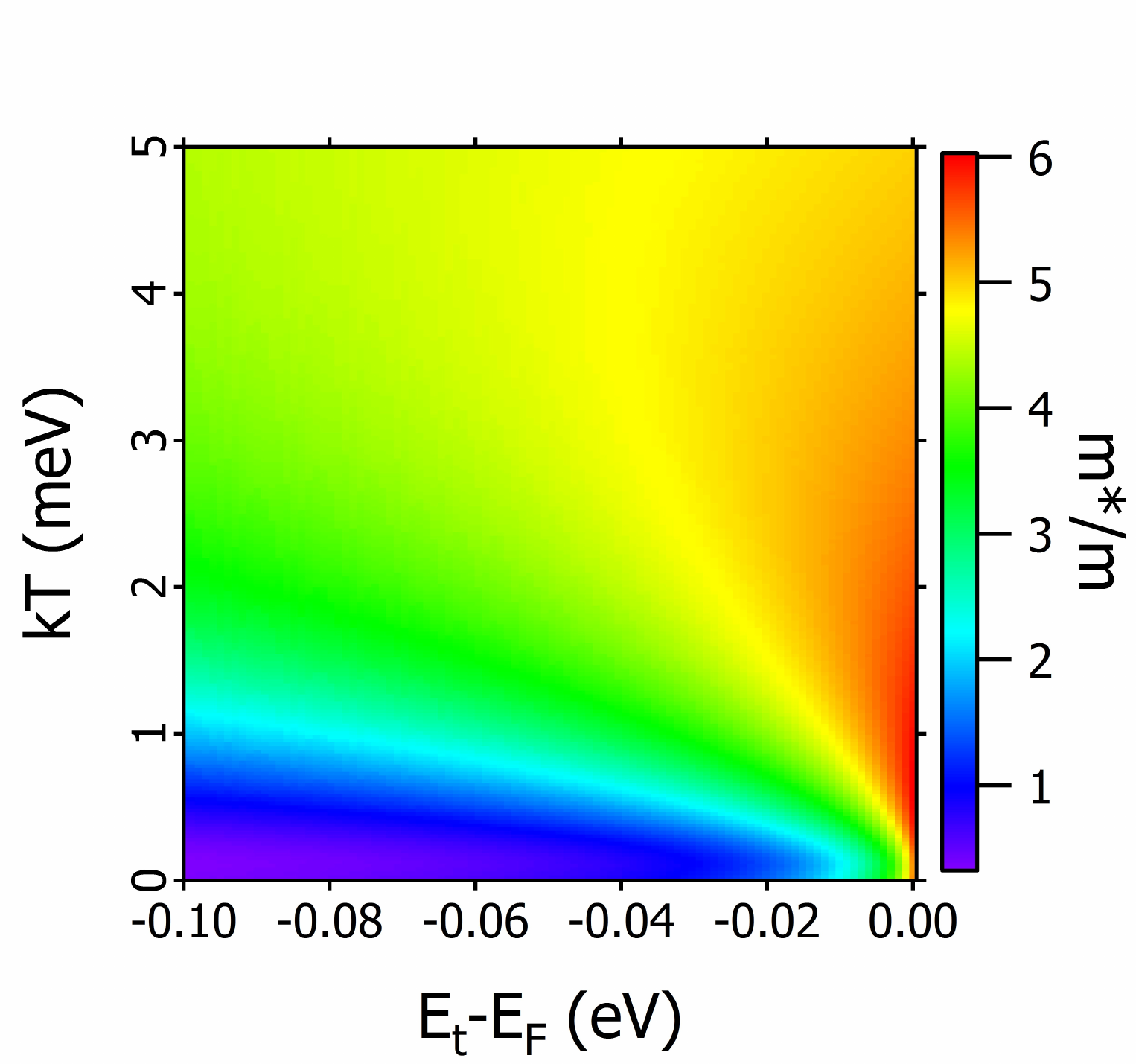}
\caption{Calculation of the effective mass as a function of the
thermal energy $k_BT$ and the shift of the top of the hole band
$E_t$. Reprint from\,\cite{Fink2016}.} \label{temp}
\end{figure}

Summarizing the results of the calculations based on the results
of the ARPES experiments, at the Lifshitz transition and only at
this transition one obtains a correlation-induced particularly
enhanced flattening of the renormalized band which is related to
an almost diverging mass at the Fermi level and at the momentum
where the Lifshitz transition occurs.  Moreover, the correlation
effects also lead to a pinning of the top of the hole pocket at
the Fermi level.

Finally a calculation of the effective mass as a function of
temperature or the thermal energy $k_{\mathrm{B}}T$ and  the shift
of the top of the hole pocket $E_t$  is presented in
Fig.\,\ref{temp}. An enhancement of the effective mass near the
Lifshitz transition and when moving away from the Lifshitz
transition a reduction at low temperatures is realized.

The presented calculations offer  a new explanation of the
NFL behavior of ferropnictides near the QCP. This
explanation is different from the traditional view which is
related to a coupling of the charge carriers to quantum
fluctuations at the end of the antiferromagnetic or nematic range. The
presented scenario is able to explain the large mass enhancement
near optimal substitution in \BFAP\ derived from the  London
penetration depth\,\cite{Hashimoto2012,Walmsley2013} and the  de
Haas-van-Alphen effect measurements\,\cite{Shishido2010} near the
QCP. Moreover, the divergent thermopower detected in \BFCA\ near
optimal doping\,\cite{Arsenijevic2013} and the thermal
properties\,\cite{Meingast2012} can be explained within the
presented scenario.

The discussion of the importance of weakly dispersing bands  at
the Fermi level for superconductivity dates back to the
explanation of high-$T_c$ superconductivity in alloys with A15
structure (e.g. V$_3$Si) by a peaked  density of states at the
Fermi level $N(E_F)$ caused by a quasi-one-dimensional single
particle electronic structure\,\cite{Labbe1966}. Later on, a
correlation of the transition temperatures with Fermi
temperatures, derived from magnetic-field penetration depth
measurements, showed that in particular the unconventional
superconductors have an effective Fermi energy close to to the
pair binding energy\,\cite{Uemura1991}. This indicates that in the
high-$T_c$ superconductors a breakdown of the Migdal
theorem\,\cite{Migdal1958} occurs  which was the basis of the
standard BCS theory of superconductivity\,\cite{Bardeen1957} and
which requires the inclusion of non-adiabatic effects and the
generalization of the Eliashberg equations\,\cite{Grimaldi1995}
and which interpolates between the BCS theory and the
Bose-Einstein condensation. Moreover, a recent compilation of the
electronic structure of unconventional
superconductors\,\cite{Borisenko2013} pointed out that most of the
high-$T_c$ superconductors have a van Hove singularity at the
Fermi level. On the other hand, many of the mentioned compounds
fulfill this criteria but have no peaked $N(E_F)$. For example the
optimally electron-doped ferropnictides have a van Hove
singularity, i.e., an edge at the chemical potential $\mu$ but all
DFT calculations predict no peaked bare band density of states at
$\mu$ in this quasi-two-dimensional electronic structure. On the
other hand the co-action of a Lifshitz transition and correlation
effects described in this review leads to a small  effective Fermi
energy and a high $N(E_F)$. This can lead to an interpolating
superconducting state between BCS and BE
condensation\,\cite{Bose1924}. The results can be generalized to
other unconventional superconductors and are possibly a recipe for
future search of high-$T_c$ superconductors.

\section{Ultrafast dynamics in Fe-122 investigated by femtosecond time-resolved ARPES}

Ultrafast pump-probe experiments investigate the response of a
material after an optical excitation by a laser pulse. On the
probed femtosecond time scales elementary scattering processes are
analyzed and provide insight into electron dynamics in the time
domain\,\cite{bovensiepen10}, complementary to the line width
analysis of static ARPES in the frequency domain. Note,
however, that the dynamics probed by ARPES and trARPES are fundamentally
different. While in static ARPES a single photo hole is analyzed,
trARPES investigates the hot quasiparticle population dynamics in
response to an optical excitation which might involve - in
particular close to the Fermi level several quasiparticle
excitations which may interact\,\cite{yang_PRL2015}. Therefore,
these two experimental approaches are highly complementary to each
other.

Femtosecond time-resolved experiments of complex materials like
charge density wave compounds and high temperature superconductors
started in the late 1990's by detecting the change of the optical
reflectivity at the Ti:sapphire laser's fundamental photon energy
of 1.5~eV and at few THz frequency\,\cite{averitt_review_2003}. At
about the same period time-resolved photoemission experiments
started to be performed in order to measure hot electron life
times several eV above the Fermi level\,\cite{nessler_prl_1998}.
Few years later relaxation of the hot electron distribution was
analyzed for \BSCCO\ by trARPES in order to determine the e-ph
coupling\,\cite{Perfetti2007} and since then this rather young
technique was developed step by step\,\cite{bovensiepen_LPR12}.

\begin{figure}[t]%
\includegraphics[width=\linewidth]{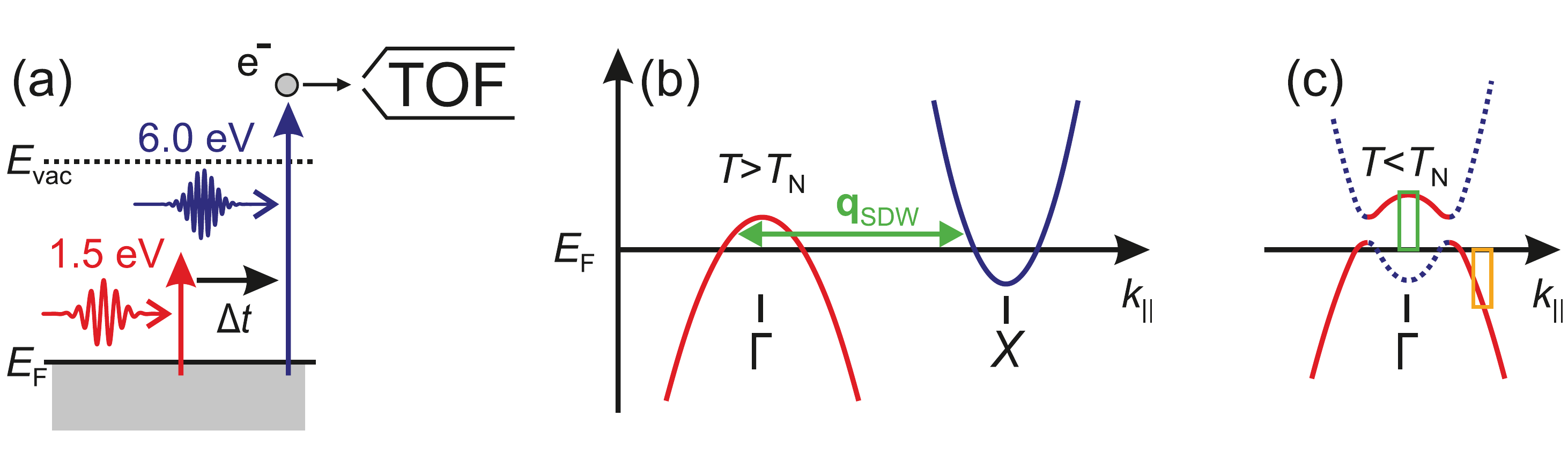}
\caption{(a) Schematic representation of femtosecond time-resolved
ARPES. Femtosecond laser pulses of photon energy 1.5 (pump) and
6.0~eV (probe) are spatially and temporally overlapped on the
Fe-pnictide surface. Emitted photoelectrons are detected by an
electron time-of flight spectrometer and analyzed as a function of
time delay $\Delta t$ between the pump and probe pulses. (b)
Simplified sketch of the Fe-122 electronic structure above the
N\'{e}el temperature $T_{\mathrm{N}}$ with hole and electron
pockets. (c) Backfolding of the electron pocket onto $\Gamma$ and
interaction with the hole pocket due to formation of
antiferromagnetic order below $T_{\mathrm{N}}$. Reprinted with
permission from\,\cite{Rettig2012}. Copyright 2012 by the American
Physical Society.} \label{trARPES_sketch}
\end{figure}

As sketched in Fig.~\ref{trARPES_sketch}(a) trARPES is essentially
a pump-probe experiment which analyzes as a function of time delay
between pump and probe laser pulses the cross correlation of the
spectral weight of optically excited quasiparticles and their subsequent relaxation
due to interaction with further fermionic and bosonic excitations.
Note that in the data reported here probing in linear
photoemission was employed, which gives rise to a time-independent
contribution to the detected photoemission intensity. The use of
laser pulses in the ultraviolet (UV) spectral range has the
advantage of a rather efficient generation processes and allows to
detect a high dynamic range of the pump-induced spectral changes.
In the present case two times the second harmonic was generated
subsequently in non-linear optical crystals which gives 6.0~eV
laser pulses for probing. The low kinetic energy of the
photoelectrons limits the accessible parallel momentum component
of the photoelectron $\hbar k_{||}$ to the center of the first
Brillouin zone close to $\Gamma$ with an extension of several 0.1
\AA$^{-1}$. Setups employing higher photon energy in the VUV or
XUV range probe the full first or further Brillouin zones but the
number of detected photoelectrons is usually lower, due to a less
efficient and more complex generation process employing higher
harmonic generation\,\cite{bovensiepen_LPR12}.

Considering the electron and hole pockets of the Fe-pnictide
electronic structure at the BZ boundary and the $\Gamma$-point,
electrons of the hole pocket or those which were scattered into
the region around the $\Gamma$-point , e.g, by $q_{\mathrm{SDW}}$
are detected, see Fig.~\ref{trARPES_sketch}(b,c). Note, that in
the present case of Fe-122 this results in a finite $k_z \approx 2 \AA^{-1}$,
corresponding to $\approx 0.6(Z-\Gamma$), at
which we detect photoelectrons. Comparing to Fig.~\ref{cokz} this
implies that we detect a part of the hole pocket in particular for
optimal doping. The results reported here analyze the response of
the electron distribution function and the chemical potential to
the pump excitation, e-ph coupling, and a spin-dependent
contribution to the relaxation dynamics which was determined in
the AFM (or SDW) ordered state.

\begin{figure}[t]%
\includegraphics[width=\linewidth]{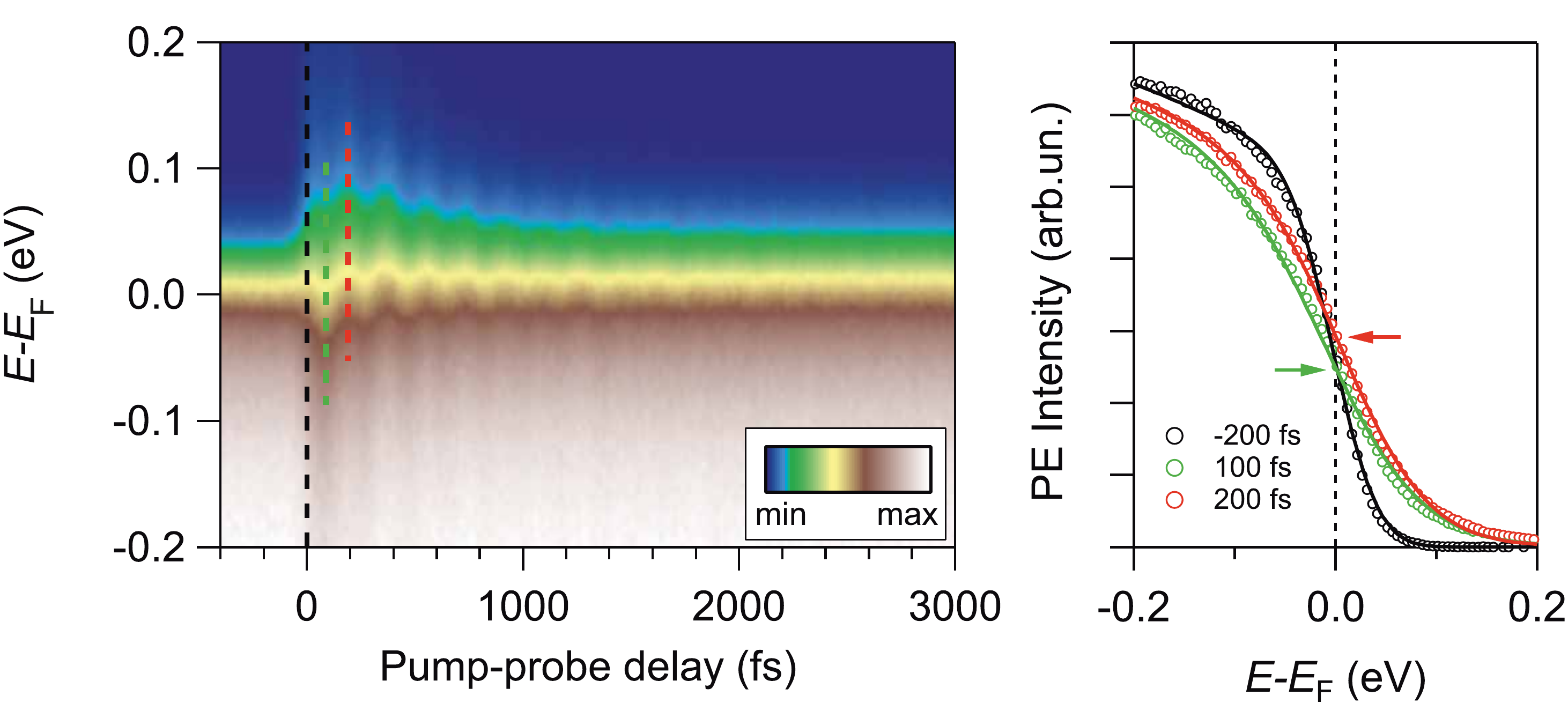}
\caption{ (a) Color coded intensity plot of the trARPES intensity
of BaFe$_{1.85}$Co$_{0.15}$As$_2$ at the $\Gamma$ point as a
function of binding energy and pump-probe delay, taken with an
incident pump fluence of $F=1.4$~mJ cm$^{-2}$ at $T=100$K. The red
and green dashed lines mark the spectra shown in (b). (b) trARPES
spectra for the first minimum (green) and maximum (red) of the
oscillation in comparison to a spectrum before excitation (black).
The solid lines are fits to the data. The Fermi level position,
$E_\mathrm{F}$, extracted from the fit is indicated by the red and
green arrows. It is oscillating due to the coherent phonon
excitation. Reproduced with permission from\,\cite{rettig_2013}.
Copyright IOP Publishing. All rights reserved.}
\label{dyn_chem_pot}
\end{figure}

\subsection{Response of the electron distribution function and the chemical potential to an optical excitation}

Absorption of the pump pulse in the investigated material results
initially in Drude and / or interband electronic excitations
exhibiting a non-equilibrium electron distribution function.
Subsequently, scattering with further electrons and bosonic
excitations changes the distribution function which eventually may
be described by a thermalized distribution function at increased
temperatures compared to the situation before pumping
\,\cite{Lisowski2004,bovensiepen_LPR12}. The material can respond to
the pump excitation in addition by a change in the electronic
structure.

Fig.~\ref{dyn_chem_pot} shows at left the time-dependent
photoelectron intensity in a false color plot around
$E_{\mathrm{F}}$. Upon optical excitation at $t=0$ the intensity
distribution broadens by about $\pm 100$~meV and exhibits a well
defined periodic variation of the high energy cutoff of the
spectrum. We conclude from this observation that the pump
excitation gives rise to incoherent and coherent excitations. The
incoherent ones relax within 1--2~ps and represent electronic
excitations. Note that although the pump photon energy is 1.5~eV
the largest pump-induced spectral changes occur at much smaller
energy which indicates that the primarily excited carriers have
scattered inelastically before being detected. The right panel of
Fig.~\ref{dyn_chem_pot} depicts three spectra taken from the data
set shown at left at the time delays indicated. The spectra after
the pump excitation at positive time delays are broadened similar
to an increased electronic temperature. We will come back to this
aspect in Sec.~\ref{e-ph_coupling} and analyze the energy content
in the excited system. The coherent spectral redistribution occurs
with a well defined frequency of 5.5~THz which can be assigned to
the Raman active A$_\mathrm{1g}$ phonon mode corresponding to a
displacement of the As atoms perpendicular to the FeAs layers. For
details of the frequency determination see
Refs.~\cite{Avigo2012,yang_PRL2014}. A detailed analysis revealed
further modes to contribute\,\cite{Avigo2012}. Why do such coherent
phonons modify the high energy cutoff of the photoemission
spectrum? Further experiments performed in collaboration with M.
Bauer, K. Rossnagel and coworkers
(Christian-Albrechts-Universit\"{a}t zu Kiel, Germany) using XUV
femtosecond pulses allowed to compare the dynamics at the
$\Gamma$- and $M$-points\,\cite{yang_PRL2014}. We observed very
similar dynamic spectral weight redistribution at the BZ boundary
and center, which corroborated the conclusion that the coherent
response originates from a periodic change of the chemical
potential, contrary to changes of a particular electronic state as
observed earlier for several other materials
\,\cite{Perfetti2006,Schmitt2011,Hellmann2012}. This effect was
explained by an inhibited restoration of the equilibrium chemical
potential after the optical excitation due to the rather small
electron diffusivity perpendicular to the FeAs-layers in
consequence of the small Fermi velocity $v_{\mathrm{F},\perp}$
such that the modulation of the chemical potential by the coherent
phonon persists until 2~ps\,\cite{yang_PRL2014}.

\subsection{Analysis of e-ph coupling}\label{e-ph_coupling}

Now we return to the incoherent dynamics and consider the
transient broadening of the electron distribution, see
Fig.~\ref{dyn_chem_pot}. An analysis of the femtosecond dynamics
by a heat bath model derived from a two-temperature model in which
the electronic excitation is accounted for by a change in the
electron temperature $T_\mathrm{e}$ and subsequent energy transfer
represented by an increase in phonon / lattice temperatures
$T_\mathrm{ph}$, $T_\mathrm{l}$ was proposed by Allen
\,\cite{Allen1987}. It was with some modification applied
successfully to various materials to describe the content of
excess energy in agreement with trARPES observations
\,\cite{Lisowski2004,Perfetti2007,bovensiepen2007}. In order to
verify this approach we compared in Ref.~\cite{rettig_2013} three
different ways to determine the electron-phonon coupling strength
$\lambda\langle \omega ^2 \rangle$.

\begin{figure}[t]%
\includegraphics[width=\linewidth]{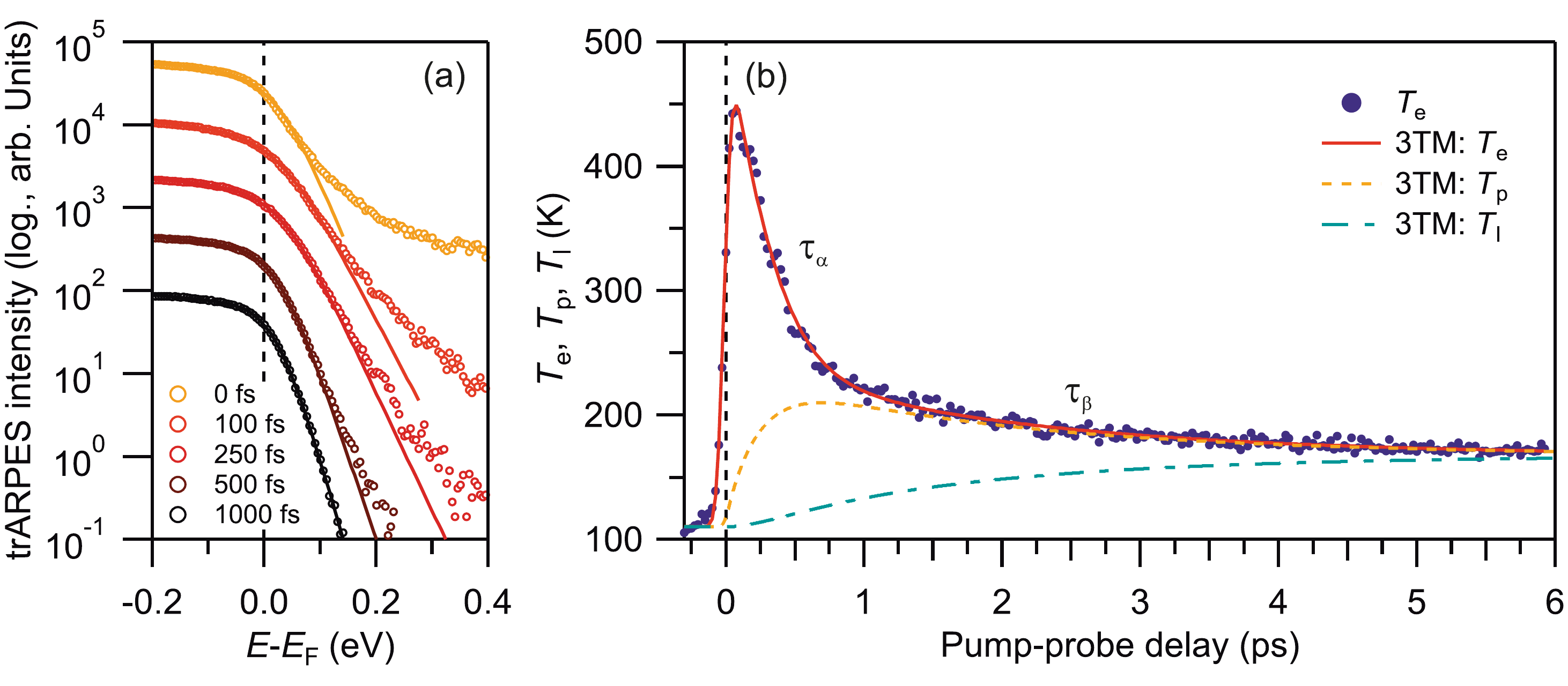}
\caption{(a) trARPES spectra of EuFe$_2$As$_2$ at $T = 100$~K and
at normal emission for various pump-probe delays on a logarithmic
intensity scale using an incident pump fluence of $F = 0.8$~mJ
cm$^{-2}$. Lines are fits to Fermi-Dirac distribution functions.
(b) Electronic temperature $T_\mathrm{e}$ determined from the
fitting shown in panel (a) and a fit to the three temperature
model. Transient temperatures of the hot phonons ($T_\mathrm{p}$)
and the rest of the lattice modes ($T_\mathrm{l}$) are shown as
dashed and dash-dotted lines, respectively. Reprinted with
permission from \cite{rettig_2013}. Copyright 2013 by IOP
Publishing and Deutsche Physikalische Gesellschaft.}
\label{elect_distr_3TM}
\end{figure}

Fig.~\ref{elect_distr_3TM} depicts in the left panel trARPES
spectra for different delays in the vicinity of $E_{\mathrm{F}}$
on a logarithmic intensity scale. With increasing pump-probe delay
the deviations between a thermalized distribution function and the
measured spectrum recedes. It takes, however, about 0.5~ps until a
thermalized electron distribution function is encountered in the
data. This might question the Allen approach conceptually on the
one hand side. On the other hand side the deviations of the
distribution function from a thermalized one occur above
100--200~meV above $E_{\mathrm{F}}$, which is above phonon
energies in the Fe-pnictides. Therefore we compare the result of
such a heat bath model with further methods to analyze energy
transfer to phonons. We start the discussion with determination of
the electron temperature by fitting it to the trARPES spectra at
different time delays and plot the result in the right panel of
Fig.~\ref{elect_distr_3TM}. The obtained time-dependent
temperature exhibits a generic behavior. The initial maximum in
temperature can be assigned to $T_{\mathrm{e}}$ and originates
from the low specific heat of the electron system. Subsequently,
the electron temperature reduces due to energy transfer to
phonons. The transient behavior of the time evolution of
temperature requires to consider two subsets of phonons, a
strongly coupled hot mode and the remaining modes which might
represent acoustic excitations. For the detailed model see
Ref.~\cite{rettig_2013}. Fig.~\ref{elect_distr_3TM} shows at right
that the overall behavior of the temperature is reproduced well by
such a three-temperature model. The energy transfer from electrons
to the hot phonons is determined by the Eliashberg e-ph coupling
function $\lambda\langle\omega^2\rangle$ times the temperature
difference between the electron and hot phonon heat baths. Since
the temperatures are known, $\lambda\langle\omega^2\rangle$ was
determined and given in the left column of table 1. Note that
there is an increasing trend for the different samples
investigated from parent to Co-doped BaFe$_2$As$_2$ to
EuFe$_2$As$_2$.

\begin{figure}[t]%
\includegraphics[width=\linewidth]{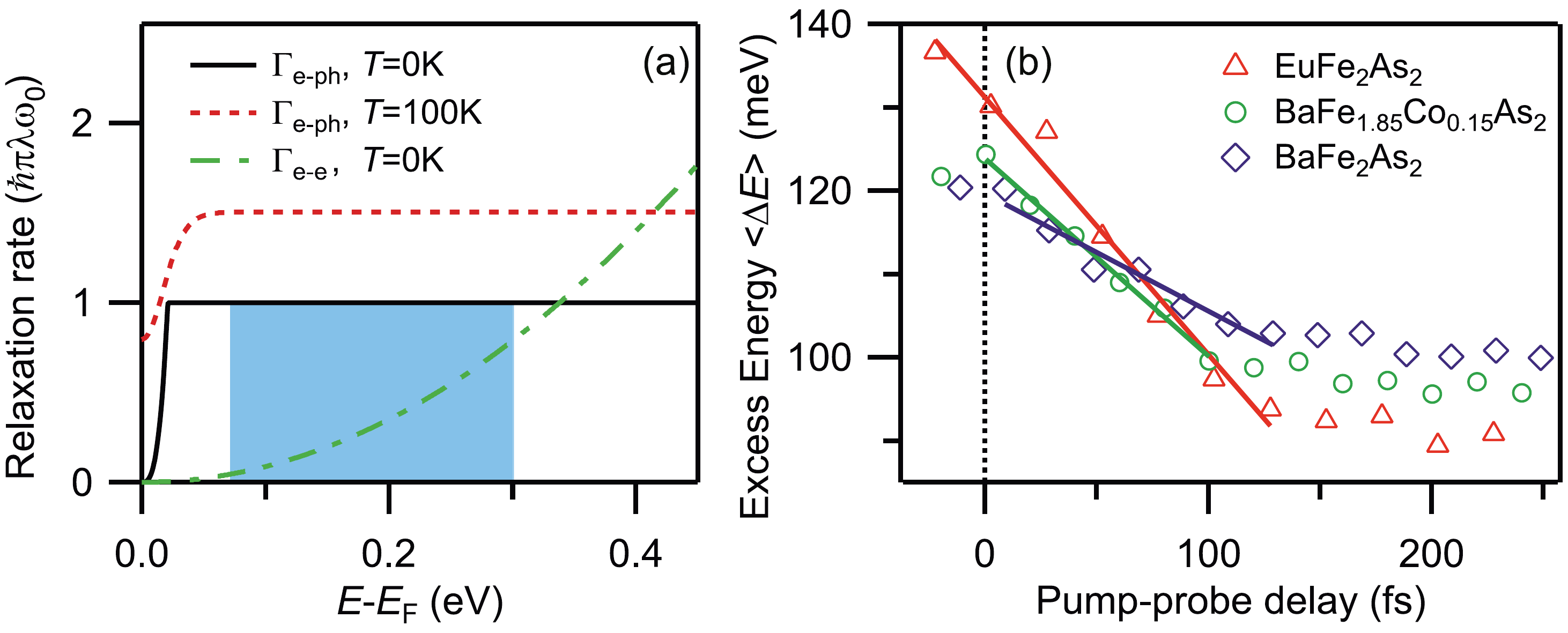}
\caption{(a) e-ph and e-e contributions to the electron decay rate
$\Gamma$. The e-ph contribution $\Gamma_{e-ph}$ calculated in the
Debye model with $\hbar \omega_\mathrm{D} = 20$~meV and $\lambda =
0.3$ at $T = 0$ and 100~K increases up to $\hbar
\omega_\mathrm{D}$ and is constant above. The electronic
contribution $\Gamma_{e-e}$ calculated for $\beta = 0.1$~eV$^{-1}$
exceeds $\Gamma_{e-ph}$ only at higher energies above 0.34~eV. The
shaded area marks the energy where e-ph scattering dominates. (b)
Mean electron excess energy extracted from trARPES data at $T =
100$K near $k_{\mathrm{F}}$ within the energy window 0.07~eV
$<\varepsilon < 0.3$~eV marked in (a). Solid lines are fits, see
text. Reprinted with permission from\,\cite{rettig_2013}. Copyright
2013 by IOP Publishing and Deutsche Physikalische Gesellschaft.}
\label{excess_energy}
\end{figure}

In a second method to determine e-ph coupling the electron excess
energy is analyzed and an assumption of a thermalized electron gas
is avoided. Thermalized and non-thermalized electrons contribute to
this energy according to their quasiparticle energy above
$E_{\mathrm{F}}$. In this analysis the decay rates of electrons is
assumed to follow the quadratic energy dependence of the Fermi
liquid theory. Very close to $E_{\mathrm{F}}$ this becomes a
negligible contribution because scattering with phonons is
dominant for an Einstein mode above its frequency $\omega_0$.
Fig.~\ref{excess_energy}(a) shows these two scattering rates in
comparison and a dominant e-ph scattering rate up to 300~meV. The
rate of energy dissipation of an electron due to emission of a
phonon with energy frequency $\omega_0$ can then be approximated
by

\begin{equation}
\label{eqn:e-ph} \frac{\mathrm{d} E}{\mathrm{d} t} =
\frac{\hbar\omega_0}{\tau}=\pi\hbar\lambda\omega_0^2\qquad
\end{equation}

and linear in time relaxation of electron energy. The rate of
energy relaxation of hot electrons has been extracted from the
experimental trARPES intensity $I(E,t)$ by analyzing the mean
excess energy by multiplying the intensity at $E,t$ with the
electron energy, for details see Ref.~\cite{rettig_2013}.
Fig.~\ref{excess_energy}(b) give the experimental results for the
mean excess energy determined within the energy interval
highlighted in blue in Fig.~\ref{excess_energy}(a). The expected
linear time dependence is indeed obtained and the slope allows to
determine $\lambda\langle\omega^2\rangle$ directly. The respective
results is added to Tab. 1, center column, and is within the error
bar in good agreement with the results obtained from the heat bath
model for all three investigated compounds.

\begin{table}
\centering \caption{Values of
$\lambda\left\langle\omega^2\right\rangle$ determined by three
different methods described in the main text.} \label{tab:1}
\begin{tabular}{l|l|l|l}
    \hline\noalign{\smallskip}
    compound & $T_e(t)$ & $\left\langle\Delta E(t)\right\rangle$ & $\tau^-$\\
    \noalign{\smallskip}\hline\noalign{\smallskip}
    \ EuFe$_2$As$_2$ & $56-65$ & $65(5)$ & $90(30)$ \\
    \ BaFe$_{1.85}$Co$_{0.15}$As$_2$ & $46-55$ & $50(3)$ & - \\
    \ BaFe$_2$As$_2$ & $30-46$ & $34(6)$ & - \\
    \noalign{\smallskip}\hline
    \end{tabular}
\end{table}

\begin{figure}[t]%
\includegraphics[width=\linewidth]{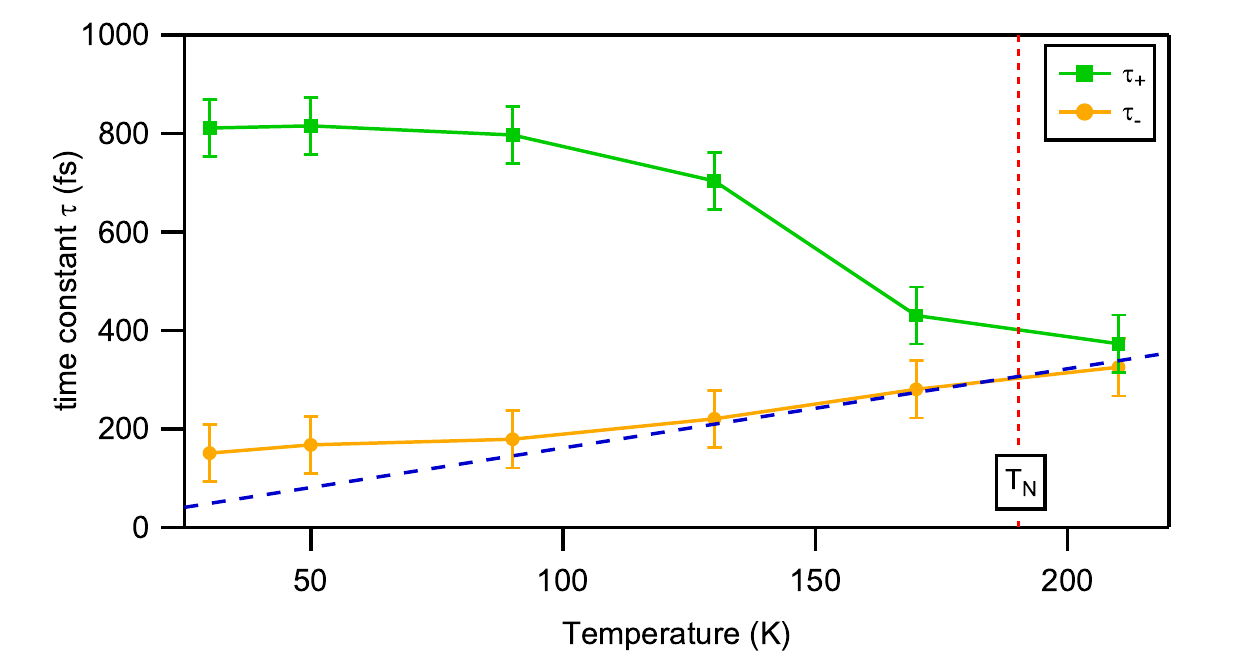}
\caption{Temperature dependent relaxation times $\tau^+$, $\tau^-$
of the spectral weight increase at $\Gamma$ and decrease at
$k_{\mathrm{F}}$, respectively,  published in
Ref.~\cite{Rettig2012} compared to a model prediction with a
linearly increasing relaxation time as a function of temperature.}
\label{Tdep_relaxation}
\end{figure}

A further way to analyze electron-phonon coupling is to analyze
the dependence of the relaxation time as a function of static
temperature. Within the limit of similar e-e and e-ph scattering
rates, which is a reasonable assumption close to $E_{\mathrm{F}}$,
see Fig.~\ref{excess_energy}(a), Kabanov and Alexandrov derived an
expression for the measured relaxation time which depends linearly
on $T / \lambda\left\langle\omega^2\right\rangle$. As is discussed
in the next section, the measured relaxation time depends on the
electron momentum and we have identified the relaxation time of
photo-excited holes $\tau^-$ to be independent on the spin density
wave formation below $T_\mathrm{N}$. The observed temperature
dependence is shown in Fig.~\ref{Tdep_relaxation} and follows a
linear temperature dependence above 70~K. Such temperature
dependent data were obtained for \EFA\ and the resulting value for
$\lambda\left\langle\omega^2\right\rangle$ agrees within the error
bar with the results of the other analysis. We therefore conclude
that the determination of
$\lambda\left\langle\omega^2\right\rangle$ is robust. These
results agree reasonably well with other experimental reports
\,\cite{Mansart2010,Stojchevska2010}.

We estimate a value of $\lambda$ for a given $\omega$. For the
Raman active A$_{\mathrm{1g}}$ mode at 23~meV which is
photo-excited, see Fig.~\ref{dyn_chem_pot}, we find
$\lambda<0.2$ for all investigated compounds. This finding is in
agreement with calculations using density functional theory, which
report values of $\lambda < 0.35$\,\cite{Boeri08,Boeri10}. The
determined $\lambda\left\langle\omega^2\right\rangle$ gives for
the lowest coupled modes of 12~meV a value of 0.5. All together
these results suggest limited importance of direct e-ph coupling
for the pairing mechanism in these materials. However, a
cooperative interaction like magneto-phonon coupling proposed due
to the dependence of the superconducting critical temperature on
the pnictogen height\,\cite{Johnston2010} could lead to an
effective enhancement of such a weak e-ph coupling. Taking into
account absolute information of the A$_{\mathrm{1g}}$ vibrational
amplitude and direction obtained in time-resolved hard x-ray
diffraction changes of the magnetic moment by 11\% were concluded
\,\cite{rettig_2015}. In case of such a cooperative coupling
mechanism e-ph coupling at such low $\lambda$ values are to be
considered as a prerequisite for superconductivity.

\subsection{Spin-dependent relaxation}

\begin{figure}[t]%
\includegraphics[width=\linewidth]{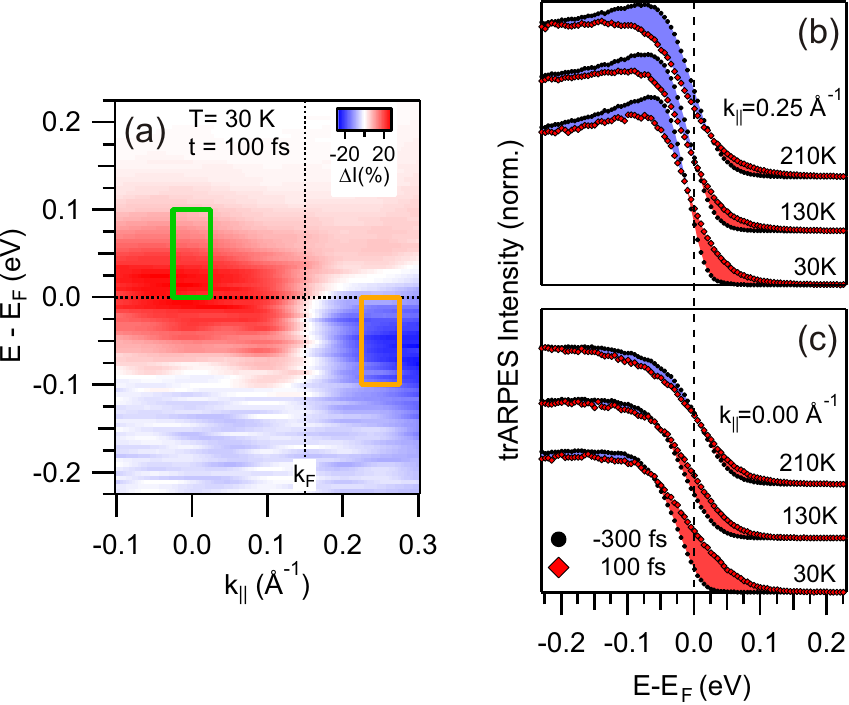}
\caption{(a) Pump-induced change of spectral weight $\delta I$ in
a false-color map for $\Delta t = 100$~fs. Blue color marks
depletion of spectral weight (hole excitations), while red marks
increased $\delta I$ (electron excitations). Boxes represent
integration areas for $\delta I$. (b) Energy distribution curves
(EDCs) for $k_{\mid\mid}=0.25$~\AA$^{-1} > k_{\mathrm{F}}$ before
(black solid circles) and $t=100$~fs after laser excitation (red
diamonds) for various temperatures. Increase and depletion of
spectral weight are marked by red and blue areas, respectively.
Spectra are vertically offset for clarity. (c) EDCs for
$k_{\mid\mid}=0$. Reprinted with permission from
\,\cite{Rettig2012}. Copyright 2012 by the American Physical
Society.} \label{dyn_spectral_weight}
\end{figure}

Besides coupling to phonons optically excited electrons can
present spin-dependent relaxation. In the strongly correlated case
of cuprates the time scales are in the range of 1-10~fs and
therefore likely below the sensitivity of the present femtosecond
photoemission experiment. However, in the antiferrimagnetically
ordered phase of parent \EFA ~we observed considerably slower
spin-dependent relaxation employing momentum-resolved detection in
trARPES. Fig.~\ref{dyn_spectral_weight}(a) depicts the
pump-induced ARPES intensity change 100~fs after the optical
excitation. In the BZ center we observe a clear intensity increase
at energies above $E_{\mathrm{F}}$, while at $k>k_{\mathrm{F}}$
the intensity decrease dominates below $E_{\mathrm{F}}$. One might
expect such a momentum-dependent intensity change due to electron
and hole excitation within the Fe 3d bands forming the hole
pockets centered at $\Gamma$. However, the analysis of the
transient spectral weight redistribution shows that such a single
electron picture only holds at $T>T_{\mathrm{N}}$. In
Figs.~\ref{dyn_spectral_weight}(b,c) spectra after optical pumping
are compared for different $k$ and $T$. At 210 K~$>T_{\mathrm{N}}$
intensity decrease and increase are observed symmetrically above
and below $E{_\mathrm{F}}$, respectively. This can be understood
in a single particle picture and an increased $T_{\mathrm{e}}$
after laser excitation, see Sec.~\ref{e-ph_coupling}. However, for
$T<T_{\mathrm{N}}$ the spectral redistribution becomes momentum
dependent. While at $k>k_{\mathrm{F}}$ the symmetric intensity
redistribution persists, at $\Gamma$ we observe a spectral weight
shift to higher energy and almost exclusively intensity increase.
Considering the backfolding and AFM energy gap in the electronic
structure in the antiferromagnetic phase, see
Fig.~\ref{trARPES_sketch}(c) allows to understand this dynamic
behavior. Upon photoexcitation, the gap closes and is filled by
electrons, evidenced by the shift of the leading edge and the
strong increase of spectral weight at $E_{\mathrm{F}}$.

The difference in dynamic spectral weight redistribution has an
interesting counterpart in the relaxation times, which were
analyzed by integrating the intensity in the energy and momentum
intervals indicated in Fig.~\ref{dyn_spectral_weight}(a) in green
and orange, see\,\cite{Rettig2012} for details. At the lowest $T$
the relaxation times at the two selected momenta differ more than
a factor 4 and the relaxation near $k_{\mathrm{F}}$ is faster than
around $\Gamma$ while they coincide above $T_{\mathrm{N}}$. In
fact, the temperature-dependent relaxation times resemble the
temperature dependence of the antiferromagnetic order parameter,
which is straightforward to understand. Relaxation near $\Gamma$
requires that the electrons relax in a spin-dependent manner such
that they match or even reestablish antiferromagnetic order. This
lowers the relaxation phase space for such electrons, which
slows down the relaxation time. Electrons near $k_{\mathrm{F}}$ do
not contribute to antiferromagnetic order and relax by energy
transfer to phonons, which we have exploited already in the
previous section in the analysis of e-ph coupling. In conclusion,
we have succeeded to determine a considerable more than four time
difference in the relaxation time of spin-dependent dynamics in
the antiferromagnetic over the paramagnetic state. Further the
influence of the structural change from tetragonal to orthorhombic
and nematic fluctuations plays a role in the detailed
time-dependent spectral weight transfer, see
Ref.~\cite{Rettig2012}, further insight might profit from
increased statistics and further optimized time-bandwidth product
of the employed UV laser pulses.

\section{Conclusion}

The collaborative activity of combining static and femtosecond
time-resolved ARPES turned out to be very fruitful in terms of
complementary information obtained in the spectral and time
domains. Essential for better understanding what is and what is
not measured by these two techniques is to appreciate the actual
observables in static and time-resolved ARPES. While the first
method probes single particle excitations, the latter one analyzes
relaxation and energy dissipation of optically excited populations
of single to many particles. In consequence, the relaxation rates
determined by these two methods can differ significantly
\,\cite{yang_PRL2015,Avigo2016}.

In the present study, both techniques indicate that the coupling of the charge carriers to phonons is considerably weaker than the coupling to electronic degrees of freedom.
In particular the results from static ARPES point to the importance of electronic correlation effects, a result which can be derived from the NFL behavior of the scattering rates. In combination with the detected Lifshitz transition near optimally doping
the results offer an explanation for the strange normal state properties, which may be different from the widely discussed quantum critical scenario. In the superconducting state, the results point to an interpolation state between BCS and BE condensation.

\begin{acknowledgement}
This work was supported by the German Research Foundation, the
DFG, through the priority program SPP 1458.
\end{acknowledgement}

\providecommand{\WileyBibTextsc}{}
\let\textsc\WileyBibTextsc
\providecommand{\othercit}{} \providecommand{\jr}[1]{#1}
\providecommand{\etal}{~et~al.}

\end{document}